\documentclass{article}

\usepackage{amssymb,amsmath}
\usepackage{fullpage}
\usepackage{graphicx}
\usepackage{appendix}
\usepackage{amsmath}	
\usepackage{slashed}
\usepackage[margin=1.0in]{geometry}
\usepackage{setspace}
\usepackage{color}
\usepackage{fancyhdr}
\usepackage{collcell}
\usepackage{datatool}
\usepackage{environ}
\usepackage{latexsym}
\usepackage{amssymb}
\usepackage{epsfig,amsmath,graphics}
\usepackage{epstopdf}
\usepackage{verbatim}
\usepackage{wasysym}
\usepackage{feynmp-auto}
\usepackage{authblk}
\usepackage{xcolor}
\usepackage{enumitem}
\usepackage[utf8]{inputenc}
\usepackage{slashed}
\usepackage{cite}

\begin{document}
\title{Redshifting the Cosmological Constant in Unimodular Gravity via Nonlinear Quantum Mechanics }







 \date{\today}

\author[1,2]{David E.~Kaplan}

\author[1]{Surjeet Rajendran}
\affil[1]{\small Department of Physics \& Astronomy, The Johns Hopkins University, Baltimore, MD  21218, USA}
\affil[2]{\small Kavli IPMU (WPI), UTIAS, The University of Tokyo, Kashiwa, Chiba 277-8583, Japan}

\maketitle

\begin{abstract}
The cosmological constant problem represents a profound conflict between quantum field theory and general relativity. Unimodular gravity offers a compelling starting point by de-gravitating the vacuum energy of the Standard Model, but this framework traditionally trades the problem of vacuum energy for a fine-tuning of initial conditions, which manifest as a ``shadow" cosmological constant. In this paper, we resolve this initial conditions problem by proposing a novel modification to gravity based on nonlinear quantum mechanics. We introduce specific state-dependent terms to the Hamiltonian, constructed from expectation values of the metric such as the average Ricci scalar. These terms alter the dynamical equations of gravity such that the shadow energy density associated with unconstrained initial conditions redshifts away with cosmic expansion, rendering it negligible at late times. The resulting cosmology is naturally dominated by matter and radiation without fine-tuning. We demonstrate that this significant infrared modification of gravity is consistent with local and cosmological tests of gravity. We comment on the possibility of testing this solution in cosmological measurements of Newton's constant. 
\end{abstract}

\tableofcontents

\section{Introduction}

The Standard Model of particle physics provides an excellent description of nature to energies at least as high as the TeV scale. The known physics of the standard model contributes to its vacuum energy density, and these contributions are at least as large as $\sim \text{TeV}^4$. However, the observational limit on the vacuum energy density of the universe constrains it to be no larger than $\sim \text{meV}^4$ \cite{SupernovaSearchTeam:1998fmf}. The extreme discrepancy between known contributions to the vacuum energy density and the observational limit on its value is the cosmological constant problem \cite{Weinberg:1988cp}. Attempts to solve the cosmological constant problem have been the source of considerable humor, rancor, and despair in the field of particle physics for the past 40 years.

There have been many attempts to solve the cosmological constant problem 
There are four strategies to solve this problem. The first is the anthropic approach \cite{Weinberg:1987dv}, where one simply accepts the existence of extreme fine tuning of certain fundamental constants as a necessary requirement for the existence of sufficiently sentient and social observers \cite{Linde:2008xf}.  The second is a symmetry-based approach (see, for example, \cite{Kaplan:2005rr}), where one posits the existence of new sectors whose vacuum energy is tied to that of the standard model via principles of symmetry in order to obtain a sufficiently precise cancellation of the standard model vacuum energy. The third is the dynamical or relaxation approach (see, for example, \cite{Abbott:1984qf,Brown:1987dd,Brown:1988kg,Graham:2017hfr, Graham:2019bfu,Kaloper:2022jpv,Kaloper:2023kua}). In this approach, the Universe is assumed to start with a large cosmological constant. Subsequently, dynamical processes over the eons of time result in the vacuum energy slowly relaxing to its observed value. The fourth approach, which will be the strategy pursued in this paper, accepts that the vacuum energy density is large but aims to suppress its gravitational and thus observational effects. The goal is to `de-gravitate' the vacuum energy (for attempts along these lines, see for example \cite{Arkani-Hamed:2000hpr, Kachru:2000hf,Sundrum:2003tb,Kaloper:2013zca}). 

De-gravitation of the vacuum energy of the Standard Model is in fact possible even within General Relativity. This approach, termed uni-modular gravity, works as follows. The metric influences the dynamics of the theory in two ways. First, the metric, its inverse and their respective derivatives, appear  as various tensor operators in the Hamiltonian density. The vacuum energy density is not coupled to these terms.  Second, a quantum field theory requires the specification of a volume element in order to define space-time integrals on the manifold on which the theory is defined. This volume element is also needed to define the Hamiltonian from the Hamiltonian density. In conventional General Relativity, this volume element (a tensor density) is also an operator -specifically, it is the square root of the determinant of the metric. The vacuum energy density does couple to this volume element. When the volume element is a non-trivial operator, the vacuum energy density gravitates. This suggests that if the volume-element was made into a trivial operator or a c-number function, then the vacuum energy density will no longer gravitate. This can be achieved even within General Relativity by quantizing the theory in the unimodular gauge - that is, the quantum theory is constructed by choosing operators so that the volume element of the theory is identically equal to one. If unimodular gravity is the true quantum theory that describes General Relativity, the vacuum energy density of the Standard Model will no longer gravitate. 

However, as is well appreciated, unimodular gravity does not solve the cosmological constant problem. The conventional description of this problem is as follows: in unimodular gravity, the constraint that the volume element is restricted to be equal to one results in the loss of one of the constraint equations of classical General Relativity. This implies that one can freely choose the initial conditions of the metric to be independent of the initial conditions of the matter in the universe. The evolution of such an unconstrained initial state can be shown to be identical to that of a universe where the constraint is obeyed, but with the matter sources being supplemented with an additional cosmological constant. Thus, in unimodular gravity, while the vacuum energy density contributions of the Standard Model no longer gravitate, a different version of the cosmological constant problem appears: why were the initial conditions of the universe such that initial state of the metric and matter exactly obeyed the classical constraint, given that this constraint is not actually a requirement of unimodular gravity?

The cosmological constant problem can be solved if the time evolution of the states that violated the constraint was such that the constraint violation redshifts ({\it i.e.} decays) as the universe expands. In this case, the vacuum energy density of the Standard Model will not gravitate and the consequences of the unconstrained initial state will decay away in time. To accomplish this, it is necessary to change the dynamical equations of gravity. This strategy needs to confront the theoretical challenges involved in modifying General Relativity and retaining the observational concordance between experiment and theory in the field of gravitation such as tests of the equivalence principle. It has recently been realized that consistent modifications of General Relativity are possible in non-linear quantum mechanics \cite{Kaplan:2021qpv}. Exploiting this fact, we will include non-linear quantum mechanical terms that modify the dynamical equations of gravity, resulting in the redshift of the unconstrained initial state of unimodular gravity. 

This paper is organized as follows. In section \ref{sec:unimodular}, we will describe the quantum theory of gravity in uni-modular gravity. In this section, we discuss why the initial state of the universe can violate the constraint equations of classical General Relativity. We will discuss how violations of constraints are a generic consequence of any quantum gauge theory. In addition, we will also discuss the ``non-gauge invariant'' aspect of this topic - why should the vacuum energy density of the Standard Model not gravitate in uni-modular gravity, when it does in fact gravitate in other ``gauge'' choices such as synchronous gauge or Newtonian gauge? We will argue that at the quantum level, these distinct ``gauge'' choices all result in a  theory of massless gravitons whose classical limit is classical General Relativity but with ``shadow'' sources of matter. The equations of state for these shadow sources depend upon the actual quantum operators that were used to define the Hamiltonian (the ``gauge'' choice). Following this discussion, in section \ref{sec:NonLinearQM}, we will show how the dynamical equations of gravity can be modified by including non-linear quantum mechanical terms into the Hamiltonian. We propose a specific term which will result in the redshift of the unimodular initial state and thus solve the cosmological constant problem. In section \ref{sec:concordance}, we show how the proposed modification is consistent with experimental tests of gravity. In this section, we discuss various cosmological scenarios such as inflation, radiation and matter domination, as well as gravitation of local sources such as the Schwarzschild and Tollman-Oppenheimer-Volkoff (TOV) solutions. We then conclude in section \ref{sec:conclusions} where we briefly comment about possible experimental tests of this scenario.

\section{Unimodular Gravity}
\label{sec:unimodular}

In this section, we review the quantization of General Relativity. We will see why the classical constraints on the initial states need not be obeyed in quantum mechanics. The time evolution of coherent ({\it i.e.} classical) initial states that violate the constraints will turn out to be that of states that obey the constraints of classical General Relativity but with additional ``shadow'' sources of matter. The equation of state of this shadow matter will depend upon the operators in the Hamiltonian. The different possible operators in the Hamiltonian would correspond to different ``gauge'' choices in the classical theory. As we will see, the quantum theory is not invariant under choice of different operators since different operator choices in fact lead to different equations of state of the shadow matter that represents the violation of the classical constraints. Nevertheless, the theory possesses sufficient symmetry under co-ordinate transformations to result in a theory of massless gravitons at the quantum level. These issues have been discussed by us in a series of papers \cite{Burns:2022fzs, Kaplan:2023fbl, Kaplan:2023wyw} - the discussion in this section is a review of these papers. We apply this understanding to the specific case of unimodular gravity. 

\subsection{Quantization of Gravity}

Our formulation is based on the following principles: 

\begin{enumerate}
    \item Quantum Field theory is the ultimate description of nature {\it i.e.} the entire universe is described by a quantum state which evolves as per the Schrodinger equation. 

    \item Classical physics, including classical  General Relativity, is simply a limit of the underlying theory. The classical limit is represented by the behavior of coherent states of the underlying quantum field theory. 

    \item In theories such as gravitation, we have experimental access to the behavior of the classical theory. Since classical physics is a limit of quantum mechanics, it is possible for several distinct quantum theories, characterized by distinct Hamiltonians, to all have the same or similar classical limit. 

    \item A limit is fundamentally a procedure that involves some loss of information. There is thus no reason to require that the underlying quantum field theory be uniquely derivable from its observed classical limit. There is thus no ``derivation'' of the quantum field theory from the observed classical physics. All that can be done is to make guesses about the underlying quantum theory (such as choices of operators). 

    \item The validity of a guess is determined by logical and experimental consistency. For gravitation, the theory should be a theory of massless gravitons and it must describe the observed time evolution of the universe. 
\end{enumerate}

Given these principles, how do we guess the correct quantum theory whose limit is classical General Relativity? The Schrodinger equation describes the time evolution of a quantum state. The quantum state is described on a spatial manifold with its evolution determined by a Hamiltonian. Accordingly, we take the space-time manifold to be form  $R\times \Sigma$ where $\Sigma$ is a spatial manifold. The metric $g$ is a quantum field that lives on this manifold and can be written as: 

\begin{equation}
g = -N dt^2 + N_i \,dt \, dx^i + g_{ij}dx^i dx^j
\end{equation}
where $t$ labels the time co-ordinate and $x_i$ are spatial co-ordinates on $\Sigma$. 

Observationally, we know that the classical evolution of gravity is well described by the Lagrangian  $\sqrt{-g} R$, with $R$ being the Ricci scalar. From this, we obtain the fact that the fields $N$ and $N_i$ lack conjugate momenta  ($\frac{\partial \mathcal {L}}{\partial \dot{g_{0\mu}}} = 0$) while the $g_{ij}$ possess non-zero conjugate momenta $\pi_{ij} = \frac{\partial \mathcal {L}}{\partial \dot{g_{ij}}}$. Since the fields $N$ and $N_i$ lack conjugate momenta, they do not represent genuine quantum degrees of freedom. To construct a Hamiltonian, these fields need to be specified. Since they are not degrees of freedom, they should be regarded as unknown but fixed parameters or functions of other dynamical operators in the theory. 

How should these operators be picked? The goal of this construction is that the classical limit of the quantum theory should yield the observed phenomenology of General Relativity. Since the classical limit is a process that loses information, one cannot uniquely derive these operators. One can make different guesses for what these operators are and see if these guesses reproduce the observed phenomenology of gravity in our world. For example, one may guess that $N = 1$ and $N_i = 0$ (corresponding to the synchronous gauge). Or, one can guess $N = 1 - \Phi_N\left(x\right)$ and $N_i = 0$ where $\Phi_N$ is the Newtonian potential operator constructed in terms of the matter fields (corresponding to the Newtonian gauge). In fact, one can guess that $N$ and $N_i$ are any of the conventional gauge ``choices'' of classical General Relativity. 

For each such guess, one can construct a canonical Hamiltonian. Each guess is a distinct quantum field theory. This is because at the quantum level, these distinct Hamiltonians cannot be mapped to each other by co-ordinate transformations. However, it can be shown that the dynamics described by any of these Hamiltonians is invariant under a co-ordinate transformation. This implies that each of these distinct quantum field theories describes a theory of massless gravitons. Moreover, for any specific coherent ({\it i.e.} classical) state, one can show that the phenomenology described by one set of guesses for $N$, $N_i$ can be mapped to the phenomenology of a different set of $N$ and $N_i$ by a co-ordinate transformation and appropriate re-definitions of the matter sources. We thus see the key point - in classical General Relativity, the invariance of the dynamics under co-ordinate redefinitions is identical to different gauge choices of $N$ and $N_i$. In the quantum theory, one can construct Hamiltonians that describe dynamics that are invariant under co-ordinate transformations. But this does not imply that the underlying quantum theory is invariant under different choices of the operators $N$ and $N_i$. The requirement that the quantum theory describe a theory of massless gravitons only requires the former condition and not the latter. 

Importantly, since these are different quantum field theories, it is not a surprise that different guesses for $N$, $N_i$ lead to differences in the gravitation of the Standard Model vacuum energy. Ultimately, the correct values of $N$, $N_i$ can only be determined with observational input. This conclusion is unsurprising when one views $N$ and $N_i$ as parameters of the quantum theory. A possible observational way to probe these parameters arises from the violations of the classical constraint equations of General Relativity. The classical limit of the quantum theory is obtained as an identity (following Ehrenfest's procedure) for quantum states that evolve as per the Schrodinger equation. Since the Schrodinger equation is a dynamical equation, it can only imply dynamical identities - it thus enforces that the classical dynamical equations of Einstein hold on expectation value. However, it cannot imply the classical constraint equations. In fact, since the Schrodinger equation is a first order differential equation, it can consistently time evolve any initial state - even states that violate the classical constraints. These violations are manifested in the classical limit as various forms of ``shadow'' stress-energy tensors. These are not new degrees of freedom, but rather correspond to the fact that the initial quantum state need not obey the classical constraints. For different choices of $N$ and $N_i$, the equation of state of this stress-tensor is different. For example, with $N = 1$ and $N_i = 0$ (the synchronous gauge), the ``shadow'' stress tensor has the equation of state of matter. But, if we instead take  $N = 1/\text{Det}\left(g_{ij}\right)$ (where the $g_{ij}$ are the spatial parts of the metric) and $N_i = 0$ (the unimodular gauge), the shadow stress energy tensor is that of the  cosmological constant. Experimental detection of these shadow stress energy tensors, while difficult, would in principle offer a way to measure these parameters. 

The fact that the equation of state of the shadow stress energy tensor in unimodular gravity is that of the cosmological constant is well known, even if it is not cast in that language. This is precisely the claim that the cosmological constant problem in unimodular gravity becomes a question of initial conditions of the universe. We will discuss this point in explicit detail in the following sub-section. But, our construction shows that the existence of states that violate constraints is a generic feature of any quantum field theory of gravitation, as long as it is governed by a single first order differential equation. The fact that equations of state of the shadow matter depends on the operators $N$ and $N_i$ is not surprising - after all, these correspond to distinct quantum field theories. Nevertheless, since each theory is invariant under co-ordinate transformations, they are all theories of massless gravitons.  With this general framework established, we now apply these principles to the specific operator choices that define the quantum theory of unimodular gravity.

\subsection{Quantization of Unimodular Gravity}

We now describe the quantum theory of unimodular gravity.  We start by writing the metric $g\left({\bf x}\right)$ in the form: 

\begin{equation}
g\left({\bf x}\right)= -dt^2 + g_{ij}\left( {\bf x}\right) dx^i dx^j
\label{eqn:MetricSynch}
\end{equation}
This metric is not in the unimodular form - but we start with this definition since it explicitly specifies  that the quantum theory is described by the dynamical degrees of freedom $g_{ij}$ and its conjugate momentum $\pi^{ij}$. This form will also be helpful when we extend the theory in the next section to include nonlinear quantum mechanical effects that will ultimately be responsible for solving the cosmological constant problem. The temporal terms $N$ and $N_i$ are parameters that have been taken to be 1 and 0, respectively. The metric and its conjugate obey the commutation relations:
\begin{equation}
[g_{ij}\left({\bf x})\right), \pi^{ab}\left({\bf x'}\right)] = i \left(\delta^{a}_{i} \delta^{b}_{j} + \delta^{a}_{j} \delta^{b}_i\right) \delta\left( {\bf x - x'}\right)
\end{equation}

The $\delta$ function in the above expression is defined by the operation: 
\begin{equation}
\int d^3 {\bf x} \, O\left({\bf x}\right) \delta\left(\bf{x - x'}\right) = O\left({\bf x'}\right) 
\end{equation}
where $O\left({\bf x}\right)$ is an operator defined on the spatial manifold $\Sigma$. 

We now define a new operator 
\begin{equation}
h\left( {\bf x}\right) =  \frac{1}{\left(-|g\left({\bf x}\right)|\right)^{1/4}} g\left( {\bf x} \right)
\label{eqn:unimetric}
\end{equation}
where $|g\left({\bf x}\right)|$ is the determinant of the metric $g\left({\bf x}\right)$. The quantum theory is still constructed from the fundamental operators $g_{ij}$ and $\pi_{ij}$. But, the Hamiltonian of the theory is constructed such that the path integral of the theory is given by the action:
\begin{equation}
\label{eqn:UGR}
S_{GR} = \int_{t_i}^{t_f} d^4 x \, \sqrt{-|h|} \left( R\left(h\right) + \mathcal{L}_M\right)
\end{equation}
where $R\left(h\right)$ is the Ricci scalar constructed from the metric $h$, $|h|$ the determinant of $h$ and $\mathcal{L}_M$ is the Lagrangian of matter fields, also constructed using $h$. Notice that $|h|$  is -1, as appropriate for unimodular gravity. This action is invariant under volume preserving spatial diffeomorphisms. This symmetry is sufficient to keep the graviton massless. It can also be checked that the only propagating modes of the metric are the spin 2 degrees of freedom and that the theory possesses a Lorentz invariant vacuum.  In fact, one can make the theory invariant under any spatial diffeomorphism by writing

\begin{equation}
h\left( {\bf x}\right) = \left( \frac{-|\eta|}{\left(-|g\left({\bf x}\right)|\right)} \right)^{1/4}g\left( {\bf x} \right)
\label{eqn:diffuni}
\end{equation}
where $\eta$ is a background volume element on the manifold $R \times \Sigma$. This makes the formulation ``background'' dependent - but there are no observational consequences of this choice that are distinct from the Hamiltonian formulation obtained from \eqref{eqn:unimetric}. Thus, for simplicity, we will proceed with \eqref{eqn:unimetric}. The classical equations of motion are obtained from $\mathcal{S}_{GR}$ by varying it with respect to the dynamical variables $g_{ij}$ and setting the variations to zero\footnote{Alternately, these can be derived by following the Ehrenfest procedure on the Hamiltonian.}. Crucially, since there are only six degrees of freedom in the $g_{ij}$, the equations of motion we get are: 

\begin{equation}
\frac{\partial S_{GR}}{\partial g_{ij}} = 0
\end{equation} 

The constraint equations $\frac{\partial S}{\partial g_{0\mu}} = 0$ do not exist. This is the explicit manifestation of the fact that since the classical equations follow from the quantum theory, the only classical equations one can obtain are those associated with the quantum degrees of freedom. The construction of the quantum theory ({\it i.e.} specification of the Hamiltonian) requires the $g_{0\mu} = \left(N, N_i\right)$ to be fixed - but, once they are fixed they can no longer be varied to yield equations of motion. This is unlike the procedure that is adopted in classical General Relativity where the variations are first performed to obtain the equations of motion with the $g_{0\mu}$ subsequently fixed. But, the axioms of classical General Relativity are irrelevant - all that matters are the equations that follow from the quantum theory. 

Let us now obtain the classical equations of motion by varying \eqref{eqn:UGR} with respect to $g_{ij}$. To do so, observe that 
\eqref{eqn:UGR} is of the form: 
\begin{equation}
S_{GR} = \int_{t_i}^{t_f} d^4 x\, \sqrt{-|h|} \left(h^{\mu \nu} R_{\mu \nu}\left(h\right) +  h^{\mu \nu}\partial_{\mu}\phi \partial_{\nu} \phi - V\left(\phi\right) + \dots \right)
\end{equation}
for various matter fields $\phi$ with $R_{\mu \nu}\left(h\right)$ constructed using the metric $h$. By construction, since $|h|=1$, it's variation $\delta |h|$ is zero.  It can be checked, using the standard arguments from classical General Relativity,  that $\delta R_{\mu \nu}$,  the variation of $R_{\mu \nu}$, is a total derivative and thus does not contribute to the equations of motion. Thus, the only non-trivial variations come from variations of $h^{\mu \nu} = \left(-|g|\right)^{\frac{1}{4}} g^{\mu \nu}$ with respect to $g_{ij}$. Thus, we have \footnote{In writing these equations, we do not explicitly write $G_N$. We are effectively in units where $8 \pi G_{N} = 1$. Alternately, the reader is reminded that the stress energy terms in the equations below implicitly contain a factor of $8 \pi G_N$. We will explicitly discuss $G_N$ when we discuss bounds on the proposed solution.}: 

\begin{equation}
\frac{\partial S_{GR}}{\partial g_{ij}} =  \frac{\partial g^{\mu \nu}}{\partial g_{ij}}\frac{\partial S_{GR}}{\partial g^{\mu \nu}} = -\frac{\sqrt{-|h|}}{\left(-|g|\right)^{1/4}} \left(g^{\mu i} g^{\nu j}\left(-|g| \right)^{1/2} \left( R_{\mu \nu} - \frac{1}{4} R h_{\mu \nu} + T_{\mu \nu} - \frac{1}{4} T h_{\mu \nu}\right)\right)
\label{eqn:StandardGR}
\end{equation}
where $T_{\mu \nu}$ is the stress energy tensor of the matter fields. Note that we have adopted Weinberg's sign  conventions in defining the Riemann tensor resulting in an overall negative sign in the RHS of \eqref{eqn:StandardGR}.  This convention will be followed throughout this paper. Setting this variation to zero, we have the equations: 

\begin{equation}
R^{ij} - \frac{1}{4} R h^{ij} = -\left( T^{ij} - \frac{1}{4}T h^{ij}\right)
\label{eqn:EinsteinUG}
\end{equation}
where the Ricci tensor and scalar are all computed with the metric $h_{\mu \nu}$. 

Let us now examine in detail how the standard model vacuum energy no longer gravitates in unimodular gravity. The vacuum energy of the Standard Model  is a contribution to $V\left(\phi\right)$. The only coupling of this term to gravity is through $\sqrt{-|h|}$. But, since the variation of this term is zero, none of the contributions in $V\left(\phi\right)$ gravitate, including the vacuum energy. The gravitational contributions of the matter sector are entirely from the kinetic terms of matter.  One may wonder if such a theory could be consistent with observations - after all, the mass of a particle enters the theory through the potential $V\left(\phi\right)$. If the potential does not gravitate, how can the mass of a particle source gravity? 

Notice that the potential does in fact affect the dynamics of gravity through the kinetic energy via the equation of motion of the matter. Specifically, the term $\frac{\partial V}{\partial \phi}$ determines the kinetic energy of the field $\phi$ and thus the potential does affect gravitation. A constant contribution to $V$, such as the cosmological constant, does not affect the equation of motion of the matter and thus this term does not affect gravitational dynamics. Even if the potential does influence gravity, one may wonder if experiment has already quantitatively  constrained the form of the stress energy tensor of matter to rule out this possibility. Notice that in unimodular gravity, the term that is missing from the stress energy tensor of matter is the term $h_{\mu \nu} \mathcal{L}_M$.

 In fact, we will now argue that in fact there is no experimental evidence that this term actually exists in conventional General Relativity when we do not restrict $-|h|$ = 1. In General Relativity, when we measure the gravitational field of an isolated body, we do so outside the body. Moreover, even if the object is made up of a number of interacting particles, the gravitational probe is a soft probe wherein the isolated object is an on-shell free particle. For a free particle that is on shell, the term $\mathcal{L}_M$ vanishes. This is easy to see since the free particle Lagrangian is bi-linear in the fields and integration by parts can be used to make  $\mathcal{L}_M$ proportional to the equations of motion. As a concrete example, consider the Klein-Gordon Lagrangian $ \sqrt{-|h|} \left(\frac{-1}{2} h_{\mu \nu} \nabla^{\mu}\phi \nabla^{\nu}\phi - \frac{1}{2}m^2 \phi^2\right)$. Upon integration by parts, this term is equal to $\sqrt{-|h|} \left(\phi/2\right) \left( h_{\mu \nu} \nabla^{\mu}\nabla^{\nu} \phi  - m^2 \phi\right)$ - that is, we see that the Lagrangian is proportional to the Klein Gordon equation, and thus vanishes on shell. This implies that even if we take a highly interacting complicated body, if we are softly measuring the gravity of this system outside the body, then $h_{\mu \nu} \mathcal{L}_M$  does not contribute to the equations of motion {\it in } General Relativity. How then does the proton gravitate in General Relativity? After all, the proton is a highly complicated object which gains mass from its QCD interactions. The potential of the proton enters the stress energy tensor through the Lagrangian term - but if these Lagrangian terms do not source gravity, how does the proton produce the appropriate gravitational field? The answer lies in the fact that when the proton is on-shell, the wave-function of the proton acquires temporal and spatial dependence that depend on the potential. The kinetic contributions are still present and it is these terms that are actually responsible for the gravitation of the proton. Concretely, a free proton is described by the  Lagrangian $i\bar{\Psi}_p \slashed{\partial} \Psi_p - m \bar{\Psi}_p \Psi_p$.  Here the mass $m$ includes all the contributions the proton gets from its interactions with QCD. These will suitably change the temporal and spatial parts of the proton's wave-function and thus gravitate via the kinetic terms. Thus both in General Relativity and in unimodular gravity, the source of gravity is entirely from kinetic terms and thus there is no difference. Interestingly, the only place where the gravitation of $h_{\mu \nu} \mathcal{L}_M$ would have shown up in General Relativity is through the gravitation of the vacuum energy - but the cosmological constant problem should make one extremely skeptical of the existence of this term. 
 
Unimodular gravity thus de-gravitates the energy density of the Standard Model vacuum. Let us now see how the cosmological constant problem re-appears in unimodular gravity as a problem of initial conditions. We will describe this problem in the quantization framework discussed in this paper. The key point is that \eqref{eqn:EinsteinUG} only contains dynamical equations. These are the only equations of motion of the theory - the constraint equations do not exist. These are thus 6 second order differential equations for the 6 metric components $g_{ij}$. These equations can always be solved for any initial conditions for $g_{ij}$ and its time derivative $\dot{g}_{ij}$. In conventional classical General Relativity, the dynamical equations are supplemented with 4 constraint equations. These are used to pin down the values of the first derivatives $\dot{g}_{ij}$. The nature of the equations are such that given any initial values $g_{ij}$ and its first derivatives $\dot{g}_{ij}$ that obey the constraint equations, the time evolution of the system can be obtained by solving the dynamical equations with these initial values. The evolution then automatically preserves the constraint equations as long as gravitation is sourced by a covariantly conserved matter stress energy tensor. But, since the classical constraint equations do not exist, one can pick arbitrary initial values for $g_{ij}$ and $\dot{g}_{ij}$ and find solutions to \eqref{eqn:EinsteinUG}. This makes sense from the perspective of quantum mechanics - in quantum mechanics, the only equation is the first order dynamical Schrodinger equation. This equation can always be solved for any initial condition. Since the classical equations are identities of quantum mechanics, it is not a surprise that one can find a solution to \eqref{eqn:EinsteinUG} for arbitrary ({\it i.e.} unconstrained) initial values of $g_{ij}$ and $\dot{g}_{ij}$. 

What is the evolution described by these unconstrained initial conditions? Given the solution to the metric, one can mathematically construct the Einstein tensor $G_{\mu \nu}$ from it. This tensor will not satisfy Einstein's equation $G_{\mu \nu} = -T_{\mu \nu}$.  Instead, using the Einstein tensor $G_{\mu \nu}$, we can mathematically construct a new tensor $T^S_{\mu \nu} = G_{\mu \nu} + T_{\mu \nu}$ so that we have: $G_{\mu \nu} = - \left(T_{\mu \nu} + T^S_{\mu \nu}\right)$. This $T^{S}_{\mu \nu}$ is the ``shadow'' stress energy tensor. Since both $G_{\mu \nu}$ and $T_{\mu \nu}$ are independently covariantly conserved, so is $T^{S}_{\mu \nu}$. This shadow tensor is a mathematical construct - it is not obtained from new physical fields. Rather, it parameterizes how the solutions obtained from unconstrained initial conditions can be interpreted in the language of classical General Relativity where one can claim that the full set (dynamical and constraint) of Einstein's equations hold but with the known matter sector supplemented by an additional shadow stress tensor $T^{S}_{\mu \nu}$.

How do we find the effective equation of state of $T^{S}_{\mu \nu}$? This can be obtained by looking at the source free solutions to the dynamical equations of the theory. This is appropriate because the dynamical equations are only sourced by the physical matter stress-energy tensor $T_{\mu \nu}$  - the shadow stress energy tensor exists only along the components that are determined by the constraints. The form of the dynamical equations (such as \eqref{eqn:EinsteinUG}) are determined by the operators in the Hamiltonian. Thus the effective equation of state of $T^{S}_{\mu \nu}$ depends on these operators. For the operators in unimodular gravity, this effective equation of state is that of the cosmological constant. But, the source of this cosmological constant is {\bf NOT} the vacuum energy density of the Standard Model - rather it represents the fact that the initial values of $g_{ij}$ and $\dot{g}_{ij}$ can be freely chosen. 

It is instructive to explicitly see this in the simple case of FRW cosmology. Accordingly, we take: 

\begin{equation}
g = -dt^2 + a\left(t\right)^2 \left(dx^2 + dy^2 + dz^2\right)
\label{eqn:FRWSynch}
\end{equation}

yielding
\begin{equation}
h = \frac{1}{a\left(t\right)^{3/2}} \left( -dt^2 + a\left(t\right)^2 \left(dx^2 + dy^2 + dz^2\right)\right)
\label{eqn:FRWUG}
\end{equation}

The dynamical equations \eqref{eqn:EinsteinUG} for a perfect fluid source with initial energy density $\rho$ and equation of state $w$ is: 
\begin{equation}
   4 a(t) \ddot{a}(t)  - \dot{a}(t)^2= -8 \rho (1 + w) a(t)^{-\frac{1}{4} - \frac{3 w}{4}}
\end{equation}
From this equation, observe that if the perfect fluid $\rho$ had the equation of state $w = -1$ (the Standard Model vacuum energy), then it would not contribute to the dynamics.  Now consider the space of solutions to this equation. This is a single second order differential equation in one unknown variable $a\left(t\right)$. Thus a solution can be obtained for any initial value of $a$ and $\dot{a}$. For any $w > -1$, let us look at the behavior of these solutions at late times. We want to know if the solution is determined by the energy density $\rho$ or by the initial conditions. This is a competition between the rate at which the energy density redshifts versus the redshift of the derivative terms of $a$. If the former redshifts more slowly than the latter, the late time solution will be determined by $\rho$. If not, the solution will be determined by the redshift of the derivative terms and thus dominated by the initial conditions. 

To determine this, start with the ansatz that at late times the solution to this differential equation is given by a power law of the form $a\left(t\right) = a_0 t^p$. Substituting, we get: 

\begin{equation}
p(-4 + 3p) (a_0 t^p)^{\frac{3(3+w)}{4}} + 8 \rho t^2 (1+w) = 0
\end{equation}

Thus, if the solution is driven by $\rho$, the expansion rate is given by $ p = \frac{8}{3(3 + w)}$. If instead the solution is driven by the initial conditions, then $\rho$ is irrelevant and the expansion rate would be given by the source-free form of the above equation, yielding $p = 4/3$. For any $w > -1$, the expansion of the derivative terms $\propto t^{3p(3 + w)/4}$ if driven by the source-free form $ p = 4/3$ is faster than the expansion $\propto t^2$ of $\rho$. This implies that $\rho$ is irrelevant for the late time behavior, which is thus entirely determined by the initial conditions. 

With the late time solution of the form $a\left(t\right) \propto t^{4/3}$, what is the form of the  inferred shadow stress energy tensor? To see this, it is useful to simply examine the behavior of the Hubble parameter. To compare the Hubble parameter to conventional cosmology, observe that the time co-ordinate in \eqref{eqn:FRWUG} depends on the scale factor and is thus not in the canonical FRW form \eqref{eqn:FRWSynch}. Relating the time $t$ in \eqref{eqn:FRWUG} to proper time $\tau$ by the relation $d\tau = dt/a\left(t\right)^{3/4}$, for the conventional Hubble parameter $\frac{da}{d\tau}\frac{1}{a\left(\tau\right)}$, we have: 
\begin{equation}
\frac{1}{a\left(\tau\right)}\frac{da}{d\tau} = \frac{1}{a\left(\tau\right)} \frac{dt}{d\tau} \frac{da}{dt} = \frac{1}{a\left(t\right)^{1/4}} \frac{da}{dt} \propto t^{-1 + \frac{3p}{4}}
\label{eqn:Hubble}
\end{equation}
for a universe that expands with power law expansion $\propto t^{p}$ in uni-modular gravity. When $p = 4/3$ as expected for the late time behavior of the solution to  \eqref{eqn:FRWUG} , we see that the Hubble parameter is constant - that is, the universe is in a de Sitter phase. The shadow stress energy tensor thus has the equation of state of the cosmological constant. 

In principle, by suitably choosing the initial conditions of $a$ and $\dot{a}$, one can arbitrarily delay the onset of this late time de Sitter phase. But, this is a fine tuning problem - why should the arbitrary initial conditions of the metric be so finely tuned in order to set the shadow stress energy tensor to zero? If the full set of  constraint equations of classical General Relativity could be imposed on the initial conditions, then the shadow stress energy would automatically be set to zero. However, the quantum theory, which is a theory of a single first order differential equation, precludes this possibility. In more conventional discussions of  unimodular gravity which are phrased in the language of classical relativity, this difficulty is phrased as the loss of a constraint equation arising from the gauge being fixed to $-|h|=1$. 

The above statement of the problem opens a new strategy to solve the cosmological constant problem in unimodular gravity. The problem can be solved if the dynamical equations of unimodular gravity are such that the late time behavior of the equations is not dominated by the initial conditions but is rather determined by the matter sources in the universe. In the language of differential equations, we want the influence of the initial condition to be a transient that dies with time. In the language of cosmology, we want the initial condition to redshift faster than the other energy densities (such as matter and radiation) so that the late time behavior of the universe is dominated by radiation and matter. Effectively, all that is required to change the coefficient of the $R h_{ij}$ term in the LHS of \eqref{eqn:EinsteinUG} to a number larger than $-1/4$ while retaining the form of the RHS. To achieve this goal, we need a consistent way to modify gravity. Such an opportunity is provided by nonlinear quantum mechanics and we will leverage that  to achieve this goal.

\section{Non-Linear Quantum Mechanics}
\label{sec:NonLinearQM}

In this section, we show how gravitation can be modified using non linear quantum mechanics, leading to corrections to the dynamical equations of gravity that can redshift the shadow stress energy tensor of unimodular gravity and thus stop it from acting like a cosmological constant. We start by reviewing the framework of nonlinear quantum mechanics and then discuss our specific proposal to solve the cosmological constant problem. We then discuss the technical naturalness aspects of our framework. 

\subsection{The Framework}

The axioms of linear quantum field theory are that the entire state of the universe is described by a quantum state $|\Psi\rangle$ whose time evolution is given by the Schrodinger equation: 

\begin{equation}
i \frac{d|\Psi\rangle}{dt} = H |\Psi\rangle
\end{equation}
Here, the quantum state $|\Psi\rangle$ is a state of various quantum fields and the Hamiltonian $H$ is an operator that is constructed from fields and their conjugate momentum operators. In linear quantum mechanics, $H$ does not depend on the quantum state $|\Psi\rangle$. It is easy to generalize the machinery of linear quantum field theory to permit nonlinear quantum mechanical evolution while maintaining causality and retaining sufficient gauge invariance to ensure that gauge bosons and gravitons remain massless. We will outline the key aspects of this construction in this paper and refer the reader to \cite{Kaplan:2021qpv} for a complete description. The main point of \cite{Kaplan:2021qpv} is that the Hamiltonian of quantum field theory can be made state dependent by including expectation values of various local operators in it. The local nature of these expectation values bakes in causality while the state dependent nature of the Hamiltonian results in non-linear time evolution. As a concrete example, consider the Yukawa theory describing a scalar field $\phi$ with a fermion $\chi$. The Hamiltonian $H_L$ of this theory is: 

\begin{equation}
H_L = \int d^3 {\bf x}  \left(H_0 + \lambda \phi\left( {\bf x}\right) \bar{\chi}\left({\bf x}\right)\chi\left( {\bf x}\right)\right)
\end{equation}
where $H_0$ is the free hamiltonian density and $\lambda$ the Yukawa coupling. The time evolution of the theory can be made non-linear by constructing the Hamiltonian: 

\begin{equation}
H_{NL} = \int d^3 {\bf x}  \left(H_0 + \lambda \phi\left( {\bf x}\right) \bar{\chi}\left({\bf x}\right)\chi\left( {\bf x}\right) + \epsilon \langle \Psi\left(t\right) | \phi\left({\bf x}\right) | \Psi \left(t\right)\rangle  \bar{\chi}\left({\bf x}\right)\chi\left( {\bf x}\right) \right)
\end{equation}
where $\epsilon$ parameterizes the size of the state dependent correction the Hamiltonian. The time evolution of this system is given by: 
\begin{equation}
i \frac{d|\Psi \left(t\right)\rangle}{dt} = H_{NL}|\Psi\left(t\right)\rangle 
\label{eqn:NLS}
\end{equation}

From the field theoretic perspective, the causal nature of this non-linearity is manifest because the non-linear term appears effectively as a local background classical field. This local background field can affect the quantum evolution of $|\Psi\rangle$ - but these are controlled by the dynamics of the quantum field theory which ensures that  fluctuations from a local source propagate within a causal cone. Alternately, it can also be shown that these expectation values, when computed on spatially well separated but entangled sub-systems $A$ and $B$, are of the form $O_A \oplus O_B$ as demanded by \cite{Polchinski:1990py} to ensure that nonlinearities remain causal. 

Non-linear quantum mechanical interactions can also be incorporated into gauge theories and gravitation. In these cases, as long as the non-linear Hamiltonian retains sufficient symmetry under c-number gauge transformations, the dynamics will continue to preserve the existence of massless gauge bosons and gravitons. For example, in the case of gravity, one can construct a non-linear Hamiltonian $H^{G}_{NL}$ of the form: 

\begin{equation}
H^{G}_{NL} = H^{G}_{L} +  \epsilon \int d^3{\bf x} \, \sqrt{-|g|} \left(\langle \Psi\left(t\right) | g_{00} | \Psi\left(t\right) \rangle  \left( \frac{\pi}{\sqrt{-|g|}} \right)^2 + \langle \Psi\left(t\right) | g_{ij} | \Psi\left(t\right) \rangle \nabla^{i}\phi \nabla^{j} \phi\right)
\end{equation}
where $\pi$ is the conjugate momentum of a scalar field $\phi$ and $H^{G}_{L}$ is the Hamiltonian of gravity coupled to $\phi$ in linear quantum mechanics. $H^{G}_{NL}$ is invariant under purely spatial co-ordinate transformations and this symmetry is sufficient to ensure masslessness of the gravitational dynamics described by it. Notice that the non-linear terms that are added to $H_{NL}$ are not universal - different fields and operators can have different levels of non-linearities. Thus, it is logically possible for example for non-linear terms to exist in the gravitational part of the Hamiltonian but be small (or even absent) in the electromagnetic sector. 

Let us now try to solve $\eqref{eqn:NLS}$ for various quantum states $|\Psi\rangle$. The state $|\Psi\rangle$ that appears in \eqref{eqn:NLS} is the full quantum state of the entire universe. This state is a massive quantum superposition and it can generically be expressed in the form $|\Psi\rangle = \alpha |U\rangle + \sum_{i} \beta_i |M_i\rangle$ where $|U\rangle$ is the quantum state of our ``world'' and the  $|M_i\rangle$ are some other ``worlds''.  This is a decohered superposition and thus in linear quantum mechanics, it is practically impossible for these worlds to influence each other. It may be that some of the $|M_i\rangle$ are nearly exactly identical to $|U\rangle$, but with some quantum states being orthogonal, resulting in decoherence. Others may be quantum mechanically very different from $|U\rangle$, but the expectation values of various operators for macroscopic entities in $|M_i\rangle$ and $|U\rangle$ may be very close to each other. Some of the $|M_i\rangle$ may have the same average FRW cosmology as $|U\rangle$, but at shorter distances $|U\rangle$ and $|M_i\rangle$ may be completely different - that is human beings or the Earth may not exist in $|M_i\rangle$ while they do in $|U\rangle$. It could even be that the $|M_i\rangle$ are wildly different from $|U\rangle$ and do not even share the average FRW cosmology with $|U\rangle$. In linear quantum mechanics, the co-efficients $\alpha$ and $\beta_i$ are irrelevant - no matter how small $\alpha$ is, or the exact nature of $|M_i\rangle$, the time evolution of the state $|U\rangle$ only cares about $|U\rangle$. This is a property of linearity - the quantum state $\alpha |U\rangle$ evolves exactly like $|U\rangle$ for any $\alpha$. Moreover, as long as the states $|M_i\rangle$ are decohered with respect to $|U\rangle$, the time evolution of $|U\rangle$ does not care about their existence in the superposition. But, this is not the case for the non-linear terms - the expectation values in \eqref{eqn:NLS} are over the entire quantum state and are thus explicitly state-dependent (thus non-linear). 

Let us now see how we can obtain the classical equations of motion of the theory. For an operator $O\left({\bf x}\right)$, we want to understand the time evolution as experienced by an observer entangled with the state $|U\rangle$ or any of the states $|M_i\rangle$ in the superposition. For this discussion, we will also consider $|U\rangle$ to be one of the $|M_i\rangle$. That is, if $|\Psi\left(0\right)\rangle = \sum_i \beta_i |M_i\left(0\right)\rangle$, we want to know the behavior of $O_i\left({\bf x}\right) = \langle M_{i}\left(t\right) | O\left({\bf x}\right) | M_{i}\left(t\right)\rangle$. Now, even though the Hamiltonian $H_{NL}$ in \eqref{eqn:NLS} is state-dependent, it is still hermitian. Thus, there is a state dependent unitary operator $U\left(t\right)_{|\Psi\left(0\right)\rangle}$ which solves \eqref{eqn:NLS}, yielding $|\Psi\left(t\right)\rangle = U\left(t\right)_{|\Psi\left(0\right)\rangle}|\Psi\left(0\right)\rangle$. Thus $|M_i\left(t\right)\rangle = U\left(t\right)_{|\Psi\left(0\right)\rangle} |M_i\left(0\right)\rangle$, yielding $O_i\left(t, {\bf x}\right) = \langle M_i\left(0\right)| U\left(t\right)^{\dagger}_{|\Psi\left(0\right)\rangle} O\left( {\bf x}\right) U\left(t\right)_{|\Psi\left(0\right)\rangle} | M_{i}\left(0\right)\rangle$. For small time intervals, we can expand $U\left(t\right)_{|\Psi\left(0\right)|\rangle}$ and obtain a differential equation for $O_i$ of the form: 

\begin{equation}
\frac{d O_i\left(t,{\bf x}\right)}{d t} = -i\langle M_i\left(t\right) | [O\left({\bf x}\right), H_{NL}(t)] | M_{i}\left(t\right)\rangle
\end{equation}
Non-zero terms arise in this equation when there are operators in $H_{NL}$ that do not commute with $O$. When $O$ is one of the canonical variables (field or conjugate momentum) obeying canonical commutation relations, these commutators effectively act as derivatives, resulting in the canonical terms in the classical Hamiltonian equations. But, note that the  nonlinear terms are of the form $\langle \Psi\left(t\right) | \Phi_1 | \Psi\left(t\right)\rangle \Phi_2\left({\bf x}\right)$ for some operators $\Phi_{1,2}$. The expectation value $\langle \Psi\left(t\right) | \Phi_1 | \Psi\left(t\right)\rangle$ always commutes with any operator. Thus, when the classical equations are being derived for some operator $O$, the term  $\langle \Psi\left(t\right) | \Phi_1 | \Psi\left(t\right)\rangle$  is not to be differentiated. If there are $N$ terms in the superposition, we thus end up with $N$ coupled equations of the form: 
\begin{equation}
\frac{d O_i\left(t,{\bf x}\right)}{d t} = F_L\left(O_i\right) + \langle \Psi\left(t, {\bf x}\right)| \Phi_1\left({\bf x}\right)| \Psi\left(t\right)\rangle \frac{\partial \Phi_{2i}\left(t,{\bf x}\right)}{\partial \Pi_O}
\end{equation}
where $\Pi_O$ is the field conjugate to $O$ and  $F_L\left(O_i\right)$ are the terms that arise from the linear parts of the Hamiltonian. That is, the system of equations has the form of $N$ independent functions $O_i$ (corresponding to the expectation values of $O$ in $N$ worlds) obeying an Ehrenfest-style relation in each world with the worlds coupled together by the expectation value $ \langle \Psi\left(t, {\bf x}\right)| \Phi_1\left({\bf x}\right)| \Psi\left(t\right)\rangle$. 

It is useful to see this result in the language of the path integral as well.  At any time $t$, this unitary operator can be fully described by its transition matrix elements in some basis. We can pick field basis states $|\phi_{ia}\rangle$, split according to different worlds $i$. In this basis, we need to specify the transition matrix elements  $\langle \phi_{ib} | U\left(T\right)_{|\Psi\left(0\right)\rangle} | \phi_{ia} \rangle $. Note that transitions from world $i$ to $j$ with $i \neq j$ are not expected to happen due to decoherence. Thus, this procedure will produce transition matrix elements that are specific to each world $i$. These elements can  be computed using the standard procedure where the time evolution is split into a large number of small time steps, with a complete set of field basis states inserted at each such step: 

\begin{eqnarray}
\langle \phi_{ib} | U\left(T\right)_{|\Psi\left(0\right)\rangle} | \phi_{ia} \rangle & = & \langle \phi_{ib} | \left(e^{-i H_{NL} \Delta t}\right)^N  |\phi_{ia}\rangle \nonumber \\
& = &  \langle \phi_{ib} | e^{-i H_{NL} \Delta t} \dots |\phi_{i,k-1}\rangle \langle \phi_{i,k-1}|e^{-i H_{NL} \Delta t} |\phi_{i,k}\rangle \dots |\phi_{ia}\rangle 
\label{eqn:NLPI}
\end{eqnarray}
Now $H_{NL}$ depends on the quantum state $|\Psi\rangle$ that is being time evolved. Since there is a guaranteed solution to \eqref{eqn:NLS} (since it is a first order differential equation), $H_{NL}$ is always known. The state dependence enters $H_{NL}$ through expectation values of the form $\langle \Psi\left(t\right) | \Phi_1 |\Psi\left(t\right) \rangle$. Following canonical methods to evaluate the differential transition matrix element $\langle \phi_{i,k-1} | e^{-i H_{NL} \delta T} |\phi_{i,k}\rangle$, we see that the full path integral of this system is of the form: 

\begin{equation}
\langle \phi_{ib} | U\left(T\right)_{|\Psi\left(0\right)\rangle} | \phi_{ia} \rangle = \int_{\phi\left(0\right) = \phi_{ia}}^{\phi\left(T\right) = \phi_{ib}} \, D\phi \, e^{i \int_{0}^{T} d^4x \left(  \mathcal{L}_0 + \langle \Psi\left(t\right) | \Phi_1 | \Psi\left(t\right) \rangle \Phi_2\right)}
\end{equation}
where $\mathcal{L}_0$ is the linear Lagrangian. Note that $\langle \Psi\left(t\right) | \Phi_1 |\Psi\left(t\right) \rangle$ is simply a known background field in this path integral - when the path integral is being performed, the paths change $\mathcal{L}_0$ and $\Phi_2$ but not $\langle \Psi\left(t\right) | \Phi_1 |\Psi\left(t\right) \rangle$. This is because the computation of the transition matrix element  $\langle \phi_{i,k-1} | e^{-i H_{NL} \delta T} |\phi_{i,k}\rangle$ yields terms that depend on the path only on terms that are non-trivial operators on the Hilbert space, unlike the c-number terms $\langle \Psi\left(t\right) | \Phi_1 |\Psi\left(t\right) \rangle$. One can obtain the classical equations of motion of this theory by demanding that \eqref{eqn:NLPI} not change when the variables of integration are changed. When the variables of integration are changed, both $\mathcal{L}_0$ and $\Phi_2$ change - but not $\langle \Psi\left(t\right) | \Phi_1 |\Psi\left(t\right) \rangle$ . Thus, there are no derivates of this term that appear in the equation of motion -  exactly as expected from the argument using Ehrenfest's relations.

Quantum mechanics has been experimentally tested now for nearly a century. No  deviations from this theory have been experimentally discovered. Given this claim, how large can nonlinearities be in quantum mechanics? One of the surprising results of \cite{Kaplan:2021qpv} (first alluded to by Polchinski \cite{Polchinski:1990py}) was that nonlinearities in quantum mechanics can be $\mathcal{O}\left(1\right)$ and yet lead to null results in laboratory tests of quantum mechanics. This is due to the fact that the non-linear terms  $\langle \Psi | \Phi_1 | \Psi\rangle$ are of the form $|\alpha|^2 \langle U | \Phi_1  | U \rangle + \sum_i |\beta_i|^2 \langle M_i | \Phi_1 | M_i \rangle$. Thus, if $|\alpha| \ll 1$ and we are in a superposition where the vast majority of the worlds $|M_i\rangle$ are very different from $|U\rangle$, then the non-linearities that can be controllably produced and detected by us (terms that only depend on $\langle U | \Phi_1 | U \rangle$) are suppressed by $|\alpha|^2$ even if the nonlinearity is fundamentally $\mathcal{O}\left(1\right)$. This is a major challenge in experimentally probing nonlinear quantum mechanics. In this limit, the non-linear terms are dominated by the expectation values $ \sum_i |\beta_i|^2 \langle M_i | \Phi_1 | M_i \rangle$. For many fields such as electromagnetism, these expectation values are generically zero and thus the effects of non-linearities in quantum mechanical evolution are dynamically zero, even if the theory is fundamentally non-zero. In this limit, the effects of non-linear quantum mechanics can only arise from fields that have non-zero expectation values in the vacuum. The two fields with such a property are the higgs and gravity\footnote{These are fields whose expectation values can affect the dynamics of relevant and marginal operators. Higher dimensional effects are possible through expectation values of operators such as $\langle G G\rangle$ where $G$ is the field strength of QCD.}. Interestingly, both of these terms are the sources of major problems in fundamental physics such as the hierarchy  and cosmological constant problems. When the expectation values are non-zero, it may be possible to exploit them to solve these outstanding problems \cite{Kaplan:2025zjj}. With this in mind,  in the next sub-section, we show how we can add non-linear terms to unimodular gravity to solve the cosmological constant problem. We subsequently show that the proposed solution is in concordance with experiment.

\subsection{The Solution}

Start with the metric $g$ defined in \eqref{eqn:MetricSynch}. Define the term $\langle g \rangle = \langle \Psi\left(t\right) | g | \Psi\left(t\right)\rangle$ and let $\langle |g|\rangle $ be  its  determinant. Form $\langle h \rangle = \langle g\rangle/ \left(-\langle |g| \rangle\right)^{1/4}$. Construct the function $\bar{R}\left(\langle h\rangle\right)$ - this is simply the expression for the Ricci-scalar constructed on the function $\langle h \rangle$. Note that $\langle g \rangle$, $\langle |g|\rangle$, $\langle h \rangle$ and $\bar{R}$ are all classical c-number functions. Given the quantum state $|\Psi\left(t\right)\rangle$ and the metric operator $g$, all of these quantities can be constructed. With these, construct the Hamiltonian: 
\begin{equation}
H_{QG} = H_{UG} + \int d^3{\bf x} \sqrt{-|h|} \left( - \kappa  \sqrt{\frac{|g|}{\langle |g|\rangle}} \bar{R}\right)
\end{equation}
where $H_{UG}$ is the Hamiltonian of unimodular gravity whose construction was discussed above. The time evolution of this system is given by the equation: 
\begin{equation}
i\frac{d|\Psi\rangle}{dt} = H_{QG} |\Psi\rangle
\label{eqn:NLQMUG}
\end{equation}
$H_{QG}$ depends on $|\Psi\rangle$ through the non-linear terms proportional to $\kappa$ and thus the time evolution is nonlinear in the quantum state. Notice that this term has the same symmetries as unimodular gravity - it is invariant under volume preserving diffeomorphisms and thus describes a theory of massless gravitons. One can make it invariant under all spatial diffeomorphisms as well by following the reasoning around \eqref{eqn:diffuni}, but at the expense of making the theory background dependent. The background dependence continues to not be an observational issue. However, for simplicity, we will stick with the formulation where we simply retain volume preserving diffeomorphisms. We note that while $\eqref{eqn:NLQMUG}$ is sufficient for red-shifting away the shadow stress energy tensor, we will include an additional term \eqref{eqn:StabilityFix} into this Hamiltonian in order to improve the stability of the dynamics of this system over a broad range of initial conditions. While that is the full solution, we start by describing the phenomenology caused by \eqref{eqn:NLQMUG} since this term is the key term that causes the red-shift of the shadow stress energy tensor. 

Let us now obtain the classical equations of motion of this system. As discussed above, the quantum state of the universe is in general a superposition of the form $|\Psi\rangle = \alpha |U\rangle + \sum_{i} \beta_i |M_i\rangle$. If there are $N$ terms in this superposition, we expect to get $N$ classical equations, one for each world $A$, with the non-linear quantum mechanical terms $\bar{R}$ and $\langle |g|\rangle$ linking these equations together. Following the procedure outlined above, for each world, we have the action:

\begin{equation}
S_{A} = \int_{t_i}^{t_f} d^4 x \sqrt{-|h|_A}\left( R_A\left(h_A\right)  + \mathcal{L}_{AM}+ \kappa \sqrt{\frac{|g_A|}{\langle |g|\rangle}} \bar{R}\right)
\end{equation}
where $h_A$, $g_A$, $R_A$ and $L_{AM}$ are quantities constructed with the fields in world A. The equations are obtained by varying $S_A$ with respect to the dynamical variable $g_{A, ij}$.  In this variational procedure, in the non-linear terms $\bar{R}$ and $\sqrt{\langle |g|\rangle}$ are not to be varied. The only part of the non-linear term that is varied is $\sqrt{-|g_A|}$. Dropping the subscript $A$ for brevity, the variational procedure yields $N$ equations for the metric $g_{ij}$ in the $N$ worlds : 
\begin{eqnarray}
\frac{\partial S_{A}}{\partial g_{ij}}  & = & \frac{\partial g^{\mu \nu}}{\partial g_{ij}} \frac{\partial S_{A}}{\partial g^{\mu \nu}}  \nonumber \\ 
&=& -\left(\frac{\sqrt{-|h|}}{\left(-|g|\right)^{1/4}} \right) \left( g^{\mu i} g^{\nu j}\left( -|g|\right)^{1/2}\right) \nonumber  \\
& \times & \left( R_{\mu \nu} - \frac{1}{4} R h_{\mu \nu} -\frac{\kappa}{2} \sqrt{\frac{|g|}{\langle |g|\rangle}} \bar{R} h_{\mu \nu} + T_{\mu \nu} - \frac{1}{4} T h_{\mu \nu}\right)= 0
\end{eqnarray}
Simplifying, we get $N$ equations of the form

\begin{equation}
R_{ij} - \frac{1}{4} R h_{i j} -\frac{\kappa}{2} \sqrt{\frac{|g|}{\langle |g|\rangle}} \bar{R} h_{i j} = -\left( T_{ij} - \frac{1}{4} T h_{ij}\right)
\label{eqn:MQMUG}
\end{equation}
where $R_{ij}$, $h_{ij}$, $g$ and $T_{ij}$ are constructed from the appropriate constituents in each world, with $\bar{R}$ and $\langle |g| \rangle$ being the non-linear elements that connect these $N$ equations. Notice that the vacuum energy of the Standard Model does not contribute to these equations - this is because we have retained the unimodular coupling of gravity to matter.  Moreover, since the underlying quantum dynamics is determined by a single first order dynamical equation  \eqref{eqn:NLQMUG}, the classical equations of motion \eqref{eqn:MQMUG} that follow from it are also dynamical equations. We do not get any constraint equations. 

Notice that these $N$ coupled equations for $N$ variables can always be solved - this follows from the fact that these coupled equations are identities that follow from \eqref{eqn:NLQMUG}, which is a single first order differential equation that can always be solved for any initial condition. Given the $N$ solutions to the metrics $g^{A}_{ij}$ for the different worlds $A$, in each world, one can calculate the Einstein tensor $G^{A}_{\mu \nu}$. In each world, this Einstein tensor will not obey Einstein's equation $G^{A}_{\mu \nu} = -T^{A}_{\mu \nu}$ in that world. An observer in that world who expects Einstein's equation to hold will interpret the dynamics as being due  to a shadow tensor $T^{A}_{S}$ in that world. 

Let us now see how to solve them. To obtain an ansatz to solve them, we need to specify the kinds of quantum states that are of interest. As discussed above, the quantum state of the universe is in general a complicated superposition of the form $|\Psi\rangle = \alpha |U\rangle + \sum_{i} \beta_i |M_i\rangle$. For the purpose of the following analysis, in the definition of $|U\rangle$, we include all quantum states in which the expectation values of all observables are narrowly spread around that of our universe. All of these states will exhibit the same effective classical behavior. We now consider the following possibilities: 

\begin{enumerate}
\item Take $\alpha = 1$. This is the case where in every ``world'' or branch of the universe, macroscopic objects have very nearly the same behavior as our world. We will call this scenario the classical scenario. 
\item Take $\alpha \ll 1$, but the $|M_i\rangle$ are such that in all of them, the average cosmological state is that of our FRW universe. In this case, the quantum state of the universe is such that the universe at small scales different branches of the wave-function have very different macroscopic properties from $|U\rangle$. For example, in most of the wave-function, the Earth may not even exist. But, in the largest scales, where the homogeneous cosmology dominates, all of the states $|M_i\rangle$ have the same homogeneous properties - that is, they all have the same matter and radiation density. This scenario is generically expected to arise from inflationary cosmology and we will call this the inflationary scenario. 
\item Take $\alpha < 1$, but with the  initial states $|M_i\rangle$ being such that their average cosmological states is  perturbatively different from the initial state of our FRW universe $|U\rangle$. That is, all the $|M_i\rangle$ have radiation and matter densities that are initially just a small departure from the energy densities in $|U\rangle$. We will call this the perturbative scenario. 
\item Take $\alpha < 1$, but with the initial states $|M_i\rangle$ wildly different from $|U\rangle$. We will call this the quantum scenario. 
\end{enumerate}

In this section, we want to focus on the evolution of the homogeneous cosmological state of the universe. Concordance with experimental measurements of local sources of matter will be discussed in the next section. We thus discuss FRW cosmology. We begin by considering the classical and inflationary scenarios. In both these cases, by assumption, the quantum state of the universe is such that the entire state has only one cosmological state. There is just one equation to solve for the scale factor   $a\left(t\right)$. Moreover $\langle |g| \rangle = |g|$ and $\bar{R} = R$. In this case, the equation \eqref{eqn:MQMUG} becomes: 

\begin{equation}
4 (6 \kappa +1) a \ddot{a}+(6 \kappa -1) \dot{a}^2+8 \rho (w+1) a^{-\frac{3 w}{4}-\frac{1}{4}} = 0
\label{eqn:fullsourceeq}
\end{equation} 

Being a second order differential equation, just like the case of unimodular gravity, it can be solved for arbitrary initial values of $a$ and $\dot{a}$. We want the late time behavior of the solutions of this equation to be determined by the matter source $\rho$ (with equation of state $w > - 1$) and not by the initial conditions. As before, we analyze the late time behavior by trying to find power law solutions of the form $a\left(t\right) = a_0 t^p$. Substituting, we get: 

\begin{equation}
p a_0^{\frac{3 (w+3)}{4}} (-24 \kappa +30 \kappa  p+3 p-4) t^{2 p}+8 \rho (w+1) t^{2-\frac{1}{4} p (3 w+1)} = 0
\label{eqn:sourceexp}
\end{equation}
If the solution is dominated by the source, we have $ p_s = \frac{8}{3 \left( 3 + w\right)}$. We need this solution to exist with $w > -1$, $a_0 > 0$ and $\rho > 0$. This is possible only when $-24 \kappa + 30 \kappa p_s + 3 p - 4 < 0$. That is, we need $ 1 + w - 2 \kappa + 6 w \kappa >0$. If the solution is driven by the initial condition, then $\rho$ is irrelevant. The expansion rate is governed by $ p_i = \frac{4 \left( 1 + 6 \kappa\right)}{3\left( 1 + 10 \kappa\right)} $. The solution will be driven by the initial condition if the kinetic terms in \eqref{eqn:fullsourceeq} (that grow like $t^{2p}$ in \eqref{eqn:sourceexp})  grow faster than the source term in \eqref{eqn:fullsourceeq} (that grows like $t^{2 - \frac{1}{4} p\left(3 w + 1\right)}$  in \eqref{eqn:sourceexp}).  Since we want the source to dominate the late time behavior, we must have $2 - \frac{1}{4} p_i\left(3 w + 1\right) > 2 p_i$. This simplifies to the requirement that $p_s > p_i$. Alternately, we need $-\frac{4}{3} \frac{\left(1 + w - 2 \kappa + 6 w \kappa\right)}{3 \left(3 + w\right) \left(1 + 10 \kappa\right)} > 0$.  Combining these conditions, if we take $\kappa <-1/10$, for all cosmologically relevant $ w > -1$, the late time behavior of the universe will be dominated by the source.  Moreover, even when the initial conditions dominate the expansion, the canonical Hubble parameter computed using the proper time (see \eqref{eqn:Hubble}) decreases as $t^{-1 + \frac{3 p_i}{4} }$. With  $ p_i = \frac{4 \left( 1 + 6 \kappa\right)}{3\left( 1 + 10 \kappa\right)} $ and $\kappa < -1/10$, this Hubble parameter decreases with time. Thus the equation of state of the shadow matter ({\it i.e.} the behavior when the initial conditions dominate the dynamics) is not that of the cosmological constant but is instead that of a energy density that steadily redshifts with the expansion of the universe. The cosmological constant problem in unimodular gravity is thus solved.  

Now consider the perturbative case. For simplicity, take the quantum state to be of the form $|\Psi\rangle = \alpha |U\rangle + \beta |M\rangle$ where the state $|M\rangle$ has initial conditions (such as the energy densities in matter and radiation, and the metric) that are perturbatively different from $|U\rangle$. The time evolution of this system is described by two metrics, $a$ and $b$, for $|U\rangle$ and $|M\rangle$, respectively. The two universes are coupled by the non-linear term $\bar{R}$ which is the classical Ricci scalar function evaluated on the classical function $\langle \Psi | g | \Psi\rangle/\langle- |g|\rangle^{1/4}$. For the FRW metric, $\langle \Psi |g|\Psi\rangle = -dt^2 + \left( |\alpha|^2 a^2\left(t\right) + |\beta|^2 b^2\left(t\right)\right)\left( dx^2 + dy^2 + dz^2\right)$ with $|\alpha|^2 + |\beta|^2 = 1$. The evolution of the system is then governed by the set of equations: 

\begin{eqnarray}
4  a \ddot{a} - \dot{a}^2+8 \rho_a (w+1) a^{-\frac{3 w}{4}-\frac{1}{4}} + \kappa F_a\left(a, b, \alpha, \beta\right) &= & 0  \nonumber \\
4  b \ddot{b} - \dot{b}^2+8 \rho_b (w+1) b^{-\frac{3 w}{4}-\frac{1}{4}} + \kappa F_b\left(a, b, \alpha, \beta\right) &= & 0 
\end{eqnarray}
where $F_{a,b}$ are the terms from \eqref{eqn:MQMUG} linking the two universes together, $\rho_{a.b}$ are the energy densities in each universe and the differential equations are to be solved with independent initial conditions for $a, b$. The two universes now interact with each other and interfere with each other's evolution. It can be checked that even if the two systems are only perturbatively different from each other, the interference leads to sharp differences in their expected time evolution - the long term behavior of the system of equations is that one of them crunches while the other grows. Thus, even though the standard model vacuum energy no longer gravitates and the redshift of the shadow tensor is not of the cosmological constant, the evolution of a superposition of states that are perturbatively close to each other is unstable. Thus for this solution to be realized in our universe, we would require the initial condition of the universe to be of the simple form of either the classical or the inflationary scenario where the quantum state that governs homogeneous cosmology is the same state across the entire superposition.

Fortunately, it is easy to make the system table. To $H_{QG}$ (in \eqref{eqn:NLQMUG}) we add the operator: 
\begin{equation}
\int d^3 {\bf  x} \, \sqrt{-h} \left( -  \bar{\kappa} \, \langle -|g| \rangle^{1/4} g^{\mu \nu} \bar{R}_{\mu \nu}\left( \langle h\rangle\right)\right)
\label{eqn:StabilityFix}
\end{equation}
where $\bar{R}_{\mu \nu}$ is the expression for the Ricci tensor constructed on the metric $\langle h\rangle$. The addition of this term modifies the classical equations of motion obeyed by the $N$ worlds to be of the form: 

\begin{equation}
R_{ij} - \frac{1}{4} R h_{i j} -\frac{\kappa}{2} \sqrt{\frac{|g|}{\langle |g|\rangle}} \bar{R} h_{i j} - \bar{\kappa} \left(\frac{\langle |g|\rangle}{|g|}\right)^{\frac{1}{4}} \bar{R}_{ij} = -\left( T_{ij} - \frac{1}{4} T h_{ij}\right)
\label{eqn:ActualFix}
\end{equation} 
It can be checked that with a slightly negative $\bar{\kappa}$, even as small as $\bar{\kappa} \sim -0.03$, perturbative superpositions remain perturbative \footnote{The smaller the value of $\bar{\kappa}$, the more tuned the initial condition needs to be.}. That is, if the quantum state is of the form $|\Psi\rangle = \alpha |U\rangle + \sum_{i} \beta_i |M_i\rangle$ with all the $|M_i\rangle$ being approximately similar to $|U\rangle$, then their time evolution is such that they all evolve with the expected equation of state. The details of this analysis are discussed in Appendix \ref{Stability}. Moreover, it is also easy to check that for these values of $\bar{\kappa}$, the dynamics of the universe is of the desired form - that is, the late time behavior is governed by the sources of energy density in the universe and that the shadow tensor redshifts away. With the inclusion of $\bar{\kappa}$, the dynamics of our universe would be expected for a larger range of initial quantum states that are perturbatively similar to our universe. The physical reason for the necessity of $\bar{\kappa}$ to stabilize the dynamics arises from the fact that when the two universes interfere,  when one of the worlds starts expanding at the expense of the other, the other world starts to contract. For perturbative stability, we need the contracting world to bounce and re-expand, at which point the expanding world starts to shrink. For homogeneuous cosmology, the former kind of a bounce requires sources that violate the null energy condition. The term $\bar{\kappa}\bar{R}_{ij}$ provides such a source term. But, this non-linear quantum mechanical term is not associated with ghost-like instabilities that are present in linear quantum mechanical field theory constructions that violate the null energy condition (see appendix \ref{Modes}).

Finally, for the quantum scenario, the behavior of the state $|U\rangle$ depends significantly on the constituents of the worlds $|M_i\rangle$. We explore this behavior in some detail in Appendix \ref{Stability}. The summary of the possible behavior is as follows. The inclusion of $\bar{\kappa}$ generally makes this system more stable - that is, if we considered two universes that were in a superposition but with each universe being governed by matter that obeyed different equations of state, the scale factors of these universes continue to grow (at least for some time). However, depending upon the exact nature of the matter in $|M_i\rangle$, the evolution of the scale factor in $|U\rangle$ could be driven by the matter in the $|M_i\rangle$ - the observer in $|U\rangle$ would attribute this expansion to various ``shadow'' stress energy tensors in $|U\rangle$. But the source of this stress energy tensor is not a new degree of freedom in $|U\rangle$ - instead it is the interference of the parallel worlds $|M_i\rangle$. Observationally, given the existence of various kinds of ``dark'' fluids in our universe such as the dark matter and the dark energy, it is unclear if there is in fact robust observational evidence that precludes this scenario. For the purposes of this paper, we will make the conservative assumption that this scenario is observationally ruled out. We will thus posit that our scenario demands that the initial state of the universe be either classical, inflationary or perturbative. With this restriction, in the next sub-section, we describe the technical naturalness of the proposed solution. 

Before concluding this section, we make a brief comment about the spatial components of the shadow stress energy tensor in this framework. Our work so far has focussed on the homogeneous component. But,  the classical spatial constraints of Einstein $G_{0i} = -T_{0i}$ also do not hold in quantum mechanics. The initial state of the metric can thus have spatial components that are unconstrained by the matter in the universe. One can solve \eqref{eqn:ActualFix} with these initial conditions as well and obtain a metric. The failure of these initial conditions will manifest itself in the existence of a shadow tensor $T^{S}_{0i}$ that will have components along the $0i$ directions. One may wonder how these components of the shadow tensor interfere with the proposed solution.  To estimate these, we can make use of the fact that $T^{S}_{0i}$ is covariantly conserved. In the limit that this tensor is initially small, we can use the metric \eqref{eqn:FRWUG} to compute the covariant conservation of $T^{S}_{0i}$ to leading order in the $0i$ components. It can be shown that these off-diagonal terms redshift like radiation and thus become sub-dominant as the universe expands. In fact, since our construction is in fact compatible with inflation (see section \ref{subsec:inflation}), an initial inflationary phase can completely wipe out $T^{S}_{0i}$ without its existence even becoming relevant for the subsequent phase of radiation domination that we had in the immediate past of our universe. 

\subsection{Technical Naturalness}

The solution proposed by us is technically natural - that is, the operators  $\propto \bar{R}$ and $\bar{R}_{\mu \nu}$ that we have written down are radiatively stable. But, there are other operators with the same symmetry structures that could have existed at the same level. We have taken their coefficients to be negligible since their presence would ruin the solution to the cosmological constant problem. While these operators are not radiatively generated, we do not have a fundamental explanation for why they are absent. Moreover, for observational reasons, the range of initial states that could lead to our observed universe is a smaller subset of quantum states due to the interference between different universes as discussed above. This restriction could be viewed as a version of ``tuning'' {\it i.e.} while it is not the formal cancellation of one parameter against another, it is nevertheless a restriction of the initial state to a smaller volume of the full Hilbert space. In this section, we comment on both of these aspects of our solution. 

The radiative stability of $\bar{R}$ and $\bar{R}_{\mu \nu}$ follows from the fact that they are classical functions ({\it i.e.} not operators) constructed from the expectation value $\langle g \rangle$. The quantum state $|\Psi\rangle$ will result in wave-function renormalization of the metric operator and thus we expect $\langle g \rangle$ to be corrected to $Z \langle g \rangle$. For a theory where the cutoff is lower than the Planck scale, the correction to $Z \ll $ - in fact, even for a Planckian cutoff, the correction is $\mathcal{O}\left(1\right)$. The factor of $Z$ multiplicatively renormalizes $\bar{R}$ and $\bar{R}_{\mu \nu}$ and are thus at most $\mathcal{O}\left(1\right)$ corrections to $\kappa$, $\bar{\kappa}$. Next, we can worry about loop processes that would involve $\kappa \sqrt{|g|/|\langle g \rangle} \bar{R}$ and $\bar{\kappa} g^{\mu \nu} \bar{R}_{\mu \nu}$. In both these cases, the operators $\langle |g| \rangle$, $\bar{R}$ and $\bar{R}_{\mu \nu}$ are classical c-number functions and not operators. Thus, they do not run in loops - rather they are simply background fields coupled to the operator $g$. The operator $g$ can run in loops and one could worry about higher order terms involving $\bar{R}$ and $\bar{R}_{\mu \nu}$ generated by such loop corrections. However, in all of these cases, since $\bar{R}$, $\langle |g| \rangle$ and $\bar{R}_{\mu \nu}$ are simply classical functions, they cannot be ``integrated out'' ({\it i.e.} there are no quantum operators in them to couple with other quantum operators to generate new terms). Thus, any new term generated by such loop processes will still contain the functions $\bar{R}$ and $\bar{R}_{\mu \nu}$. These are thus higher dimensional operators and thus have negligible impact on the physics. This establishes the radiatively stability of these operators. Notice however that we could have written down additional terms in the Hamiltonian of the form $\langle R \rangle$ and $g^{\mu \nu} \langle R_{\mu \nu} \rangle$ - that is, instead of constructing the Ricci scalar and tensor from the expectation value of the metric, we could have directly written down the expectation value of the Ricci scalar and tensor. These terms would have the same symmetry properties as $\bar{R}$ and $\bar{R}_{\mu \nu}$. But, if either of them had coefficients that were of the same size as $\kappa$, $\bar{\kappa}$, it would lead to a less severe, but nevertheless, still deadly, cosmological constant problem. That is because $\langle R \rangle $ and $\langle R_{\mu \nu} \rangle$ get contributions from gravitational contributions to the cosmological constant. If the cutoff of the Standard Model is $\Lambda \sim \text{TeV}$, these terms are $\sim \Lambda^6/M_{pl}^2 \sim \left( \text{10 keV}\right)^4$. The coefficients of these terms thus have to be appropriately small in order to not re-generate a cosmological constant. While we do not have a fundamental explanation for why these terms are small, we note that this structure is radiatively stable. The linear quantum mechanical operators in \eqref{eqn:NLQMUG} cannot generate any non-linear quantum mechanical term - that is because the non-linear terms fundamentally involve the expectation values of field operators in the quantum state that is being time evolved. Since the linear terms do not contain such terms, non-linearity is not radiatively generated by linear quantum mechanics. The only other source of such a term in this theory would be through loops involving $\bar{R}$ and $\bar{R}_{\mu \nu}$ - but as argued above, since these are classical functions, they are simply background fields whose structure cannot be altered by loops. One possible avenue to explore in ensuring that the UV only produces terms such as $\bar{R}$ and $\bar{R}_{\mu \nu}$ might be to posit that the non-linear quantum mechanical terms in the UV all arise from expectation values of the metric $\langle g \rangle$. In this case, the effective non-linear terms that emerge in the infra-red would all be functions of $\langle g \rangle$ like $\bar{R}$ and $\bar{R}_{\mu \nu}$ but not terms such as $\langle R\rangle$ and $\langle R_{\mu \nu}\rangle$. 

Next we discuss the restriction on the initial states that could lead to our observed universe. In both unimodular gravity and our proposed nonlinear extension, while the vacuum energy of the Standard Model does not gravitate, we need to pick a smaller set of initial quantum states in order to reproduce our observed universe than possible in the Hilbert space. In unimodular gravity, the initial conditions of that quantum state lead to a shadow stress energy tensor that behaved like a cosmological constant and thus this shadow stress energy tensor had to be extremely fine tuned with the matter energy density in the universe in order to reproduce our observed world. Since this is an initial condition, unimodular gravity by itself does not provide a mechanism to achieve such a tuning. In our case, the shadow stress energy tensor redshifts away and thus, we do not have a cosmological constant to contend with. But, interference between different universes that could exist in the superposition could lead to behavior that is observationally different from what is empirically observed. It is this empirical fact that restricts the initial condition. 

The nature of the restriction on the initial state in our scenario is significantly different from the case of unimodular gravity. First, with $\bar{\kappa}$ turned on, the universe can be in a perturbative superposition (in addition to the classical and inflationary case). This is a much bigger class of initial states than the fine tuned initial states needed in unimodular gravity. Moreover, any perturbative superposition with any initial value of the matter and shadow energy density would evolve to a FRW universe without a cosmological constant. The restriction is thus really on ``wild'' superpositions where the parallel universes are dramatically different from our universe. This allows for a larger class of quantum states to evolve into our universe, greatly ameliorating the cosmological constant problem. Moreover, unlike unimodular gravity where there is no mechanism to control the initial condition, non-linear quantum mechanics may provide a way to tackle this problem. In linear quantum mechanics, the evolution of an initial  de-coherent superposition of the form $|\Psi\rangle = \alpha |U\rangle + \sum_{i} \beta_i |M_i\rangle$ is such that each universe evolves independently of the others, with no mechanism to exchange energy between them. But, this is not the case in non-linear quantum mechanics. One of the most robust aspects of any causal non-linear modification of quantum mechanics is the ability for different worlds to influence each other even in the presence of decoherence \cite{Kaplan:2021qpv,Polchinski:1990py}. In fact, this is exactly the mechanism that is responsible for creating interference between different universes (through the $\kappa$ term). But, as we have seen (through the $\bar{\kappa}$ term), non-linear interactions can in fact prevent the different branches of the universe from spreading away from each other. If non-linear terms are added to other fields in the theory, it may be possible to exchange energy across different branches of the wave-function so that they come into equilibrium and thus achieve the same average (or even perturbatively similar) cosmological state. If this occurs at a rate faster than gravity, it might be possible for wildly different initial states to approach the required class of states needed to reproduce our observed universe. We have not attempted to construct such a model but the fact that such a possibility can exist in principle distinguishes this solution from the case of unimodular gravity where there is no known plausible way to achieve the required tuning of the initial conditions. 

\section{Concordance with Observations}
\label{sec:concordance}
The modifications proposed in \eqref{eqn:ActualFix} are $\mathcal{O}\left(1\right)$ corrections to General Relativity. In this section, we show that despite extensive tests of gravity, the proposed scenario is presently experimentally viable. Interestingly, it might be possible to experimentally probe this scenario with future experiments, as we briefly discuss in our conclusions (section \ref{sec:conclusions}). Experimental probes of gravity arise from cosmology through the study of the cosmic microwave background and big bang nucleosynthesis that have helped us pin down the eras of matter and radiation domination. Local tests of gravity involve precision measurement of the motion of test bodies and through it, the establishment of the Schwarzschild/Kerr metric as the solution to gravitating bodies. Recently, these tests have been performed even in the regime of strong field gravity through observations of black hole mergers in gravitational wave detectors. One of the simple reasons why our proposed modification is ill constrained by these measurements is the fact that the modification simply alters the metric created by a source. It does not change how matter couples to the metric. 
The theory thus obeys the equivalence principle and thus precision tests of equivalence principle do not constrain this theory. The theory will be in agreement with experiment as long as the metric produced by this theory is in agreement with the metric that is expected to be sourced by the matter, within experimental uncertainties. We discuss these for radiation and matter domination in sub-section \ref{subsec:cosmo} and for local sources of matter in sub-section \ref{subsec:local}. As with all tests of non-linear quantum mechanics, the applicability of a test depends upon the unknown initial state of the universe. If $\alpha \sim \mathcal{O}\left(1\right)$, the scenario can be tested in the laboratory by controllably sourcing and detecting the non-linearity. If $\alpha \ll 1$, controlled laboratory tests are difficult - one has to rely on cosmological measurements. For each of these categories, we will thus discuss two cases: $\alpha \sim \mathcal{O}\left(1\right)$ and $\alpha \ll 1$. 

Finally, in this section, we will also discuss concordance of this scenario with cosmic inflation. Inflation is by no means an empirically verified fact of nature. Nevertheless, the inflationary framework provides a good explanation for the observed spectrum of density fluctuations in the universe. It is also a natural reason why $\alpha \ll 1$, leading to the observed linearity of quantum mechanics in local experiments. Inflationary cosmology, in concert with the proposed non-linear modifications, may thus provide a natural explanation for why we see a world where quantum mechanics is observed to be linear but with a non-gravitating standard model vacuum energy. At first glance, these ideas seem to be at odds with each other - inflation would seem to require the gravitation of the vacuum energy while the proposed mechanism explicitly de-gravitates vacuum energy. How could inflation be consistent with this model? We address this interesting issue in sub-section \ref{subsec:inflation}. 

\subsection{Radiation and Matter Domination}
\label{subsec:cosmo}
We choose $\kappa < -1/10$, $\bar{\kappa} <0 $ so that the late time dynamics of the universe is dominated by the radiation and matter in the universe, with the dynamics remaining stable when the universe $|U\rangle$ is perturbed by interference with other universes that are perturbatively similar to $|U\rangle$ . In the case of radiation domination, it can be verified that $R = \bar{R} = 0$, leaving $\bar{R}_{ij} \approxeq R_{ij}$. Thus, $\kappa$ does not affect the behavior of the universe during radiation domination. The observational bound on $\bar{\kappa}$ depends on the cosmological scenario.

Interestingly, in the case of the classical scenario, there are no bounds on $\bar{\kappa}$. This is because the cosmological and local values of $G_N$ would be equal in this scenario for all classical sources of matter. One might think that this scenario could be tested by placing matter in a quantum superposition as in the case of \cite{Page:1981aj}. But, when matter is placed in a quantum superposition, the metric outside the matter distribution in each arm of the superposition is still the Schwarzschild metric. For this metric, at the linearized level in the gravitational potential, $\bar{R}_{ij}$ is zero. Thus the term $\bar{\kappa}\bar{R}_{ij}$ does not source corrections to the equations of gravity. 

In the inflationary scenario, $\bar{\kappa}$ is degenerate with the value of the gravitational constant $G_N \rightarrow G_{N}\left( 1 +  \bar{\kappa}\right)$. The existence of a non-zero $\bar{\kappa}$ implies that cosmological measurements of $G_N$ would differ from laboratory measurements of $G_N$. Current cosmological measurements imply $|\bar{\kappa} |\lessapprox 2 \times 10^{-2}$ \cite{Lamine:2024xno}, although given the Hubble tension, this number could be as big as $\lessapprox 0.05$ \cite{Benevento_2022}. In fact, there is no model independent bound on $\bar{\kappa} < 0$. This is because when $\bar{\kappa} < 0$, the value of $G_N$ relevant for the early universe is smaller than the value of $G_N$ inferred from laboratory measurements. But, this can, in principle, be compensated by additional dark radiation. For our purposes, we will explore the phenomenology for $\bar{\kappa}$ in the range $-2.5 \times 10^{-2} \leq \bar{\kappa} \leq -10^{-2}$.

During matter domination, the universe continues to expand as expected in FRW cosmology. However, for fixed initial values of the matter density and initial scale factor, the proportionality constant $a_m$  that governs the late time solution\footnote{Note that we are in unimodular time, with the metric anstaz \eqref{eqn:FRWUG}} $a\left(t\right) = a_m t^{8/3}$ is now a function of $\kappa$ and $\bar{\kappa}$. $a_m$ can be constrained using observations of the red-shift at matter-radiation equality or recombination (or really, any other known event during matter domination) through the following.  In our universe, we know the that the universe was radiation dominated during big bang nucleosynthesis. We also know the temperature when this process occurred. During radiation domination, the universe expands as $a\left(t\right) = a_r t^{4/5}$. Given known radiation density $\rho_r$ and  matter density $\rho_m$ at this known temperature, the temperature where matter-radiation equality (or recombination) must occur is a prediction of the theory. This temperature is dynamically governed by the change in the scale factor of the universe between nucleosynthesis and matter-radiation equality and is thus a function of $a_r$ and $a_m$. In this theory, $a_m$ is a function of $\kappa$ and $\bar{\kappa}$.  Using the observed value of the temperature at matter-radiation equality and the other energy densities in the universe, one can, in principle, constrain $\kappa$, $\bar{\kappa}$. However, precise quantitative comparisons with $\kappa$ are difficult to do since the dominant expansion of our universe during matter domination is driven by an known ``dark matter'' component whose nature is completely unknown. The known baryonic physics is not strong enough to impose meaningful limits on $\kappa$. 

Let us now see this argument in some detail. For the purpose of presentation, we show how $a_m$ depends on $\kappa$, neglecting the small correction to this value from the non-zero value of $|\bar{\kappa}| \lessapprox 2 \times 10^{-2}$. In this limit, $a_m \propto \rho_{m}^{4/9}/ \left( 1 - 8 \kappa  + 24 \kappa^2  -32 \kappa^3  + 16 \kappa^4\right)^{1/9}$. To test observational concordance,  we can  compare the values of $a_m$ when $\kappa = 0$ (the null hypothesis, where we  assume $\Lambda$CDM to be true) to the value of $a_m$ with $\kappa < 0$. To get the standard model expectation, we set $\kappa = 0$ in this expression but observe that in $\Lambda$CDM, the correct cosmology requires us to take $\rho_m \approxeq 6 \rho_b$ where $\rho_b$ is the known baryonic density. Thus, the era of matter radiation equality occurs for a value of $a_0$ that is larger than expected from the known baryon density $\rho_b$. With $\kappa < 0$, observe that the denominator of $a_m$ becomes bigger than 1 - thus to maintain the same value of $a_m$ as expected for $\Lambda$CDM, we need to simply increase $\rho_m$. In other words, with $\kappa < 0$, the term $\kappa \bar{R}$ effectively contributes like negative energy matter during matter domination and thus quantitative agreement with $\Lambda$CDM can be obtained by appropriately increasing the unknown amount of dark matter in the universe. There is thus no actual observational limit on this parameter from measurements of homogeneous cosmology. 

At the inhomogeneuous level, it might be possible to constrain large negative values of $\kappa$ using the spectrum of density fluctuations.  If $\kappa \ll -1$, a large positive energy density in the dark matter would be necessary to fix matter-radiation equality at the observed redshift. If the dark matter density is large and if $\delta \rho/\rho$ in the dark matter is still $\sim 10^{-5}$, it could lead to large inhomogeneous gravitational potentials during the cosmic microwave background. But, there is no model independent limit on $\kappa$ since physics in the dark sector can suppress the growth of such perturbations (if the dominant dark matter species is `warm', for example). 
For our purposes, we note that our solution works even with $\kappa \sim -1/5$, which implies a correction to the dark matter density of around 40\%.

When $\alpha \approxeq 1$ (the classical case), $\bar{R}_{ij} = R_{ij}$ and $\bar{\kappa}$ is simply a renormalization of the overall value of $G_N$. Deviations from this result are possible if the matter was placed in a quantum superposition. But, as we will see below, the effects of $\bar{\kappa}$ in linearized gravity are negligible. Since we do not have access to strongly gravitating objects that we can controllably place in a superposition, this parameter is unconstrained. Similarly,  local experiments do not constrain $\kappa$ since $R = \bar{R} = 0$ in these. The only constraint on $\kappa$ term comes from cosmology, where, as per the above analysis, $\kappa$ contributes like a negative energy component of dark matter and is degenerate with the unknown physics responsible for dark matter. 

\subsection{Schwarzschild and TOV Solution} 
\label{subsec:local}
If $\alpha \ll 1$, the Schwarzschild and TOV solutions in this theory are practically unaltered from that of General Relativity. This is because in this limit, the local sources of matter that are turned on to create gravity constitute a negligible change to $\langle h \rangle$ and thus a negligible change to $\bar{R}$ and $\bar{R}_{ij}$. The average metric $\langle h \rangle$ is dominated by the homogeneous FRW metric and thus $\bar{R}$ and $\bar{R}_{ij}$ are both $\propto H^2$ where $H$ is the Hubble constant today, utterly negligible in any local test of gravity. 

Let us study the Schwarzschild solution when $\alpha \approxeq 1$. In this limit $R_{ij} = \bar{R}_{ij}$, $R = \bar{R}$ and $g = \langle g \rangle$. We want vacuum solutions to \eqref{eqn:ActualFix}, with the boundary condition that we have a spherically symmetric point source. To obtain these, start with the metric ansatz: 

\begin{equation}
g = -dt^2 + h_0\left(r\right) dr^2 + f_0\left(r\right) r^2 d\Omega^2
\end{equation}
and obtain the unimodular metric: 
\begin{equation}
h = \frac{1}{h_0\left(r\right)^{1/4}f_0\left(r\right)^{1/2}} \left(  -dt^2 + h_0\left(r\right) dr^2 + f_0\left(r\right) r^2 d\Omega^2\right)
\label{eqn:UniSchwarz}
\end{equation}
We then want to find vacuum solutions to the equation: 
\begin{equation}
R_{ij}\left(1 - \bar{\kappa}\right) - \left(\frac{1}{4} +  \frac{\kappa}{2}\right) R h_{ij} = 0 
\label{eqn:kappaSchwarz}
\end{equation}
Dividing by $\left(1 - \bar{\kappa}\right)$, we see that while this term can affect the normalization of the source of the Schwarzschild solution (a point we will get to later in this sub-section), its effects on the analytic structure of the  Schwarzschild solution is degenerate with $\kappa$.  First, note that the Schwarzschild solution $R_{ij} = 0$ with $R = 0$ is still a solution to \eqref{eqn:ActualFix}. We want to show that it is in fact the only solution.

 We offer two arguments for this - the first one, which employs  perturbativity and analyticity, is a general argument that can also be extended to the Kerr solution. The second is a more explicit computational argument. For the perturbative argument, imagine that $\kappa \ll 1$ so that we can reasonably calculate corrections to Schwarzschild using perturbation theory. To obtain the $\mathcal{O}\left(\kappa\right)$ correction to Schwarzschild in \eqref{eqn:ActualFix}, we first solve  \eqref{eqn:ActualFix} by setting $\kappa = 0$. The solution thus obtained, which is in fact the Schwarzschild solution, is used to calculate $R$ to zeroth order in $\kappa$. This is then plugged into $\eqref{eqn:ActualFix}$ to obtain the $\mathcal{O}\left(\kappa\right)$ correction to Schwarzschild. But when $\kappa = 0$, the Schwarzschild solution is such that $R= 0$. Thus there is no $\mathcal{O}\left(\kappa\right)$ term in \eqref{eqn:ActualFix} to source a correction to Schwarzschild at $\mathcal{O}\left(\kappa\right)$ and the solution continues to be the Schwarzschild solution. One can now compute the  $\mathcal{O}\left(\kappa^2\right)$ correction using the same argument and realize that this also vanishes, thus proving that there are no analytic corrections to Schwarzschild at any order in $\kappa$. Since the system does not permit for the existence of new modes  (see appendix \ref{Modes}) that could potentially lead to non-analytic behavior, we see that there cannot be any correction to Schwarzschild. The same logic can also be applied to the Kerr solution, which is also a unique vacuum solution (with the assumed symmetries of the problem). Similarly, effects of General Relativity such as the Lense-Thirring effect that are obtained from boosts of the Schwarzschild metric are also unaffected. 
 
A more computationally explicit argument can also be given for the Schwarzschild case. Starting with \eqref{eqn:UniSchwarz}, use the fact that the theory is invariant under spatial coordinate transformations (see discussion around \eqref{eqn:diffuni}) to redefine the radial coordinate $r$ so that the spherical parts of the metric can be put in the canonical form. The resulting metric is then of the form: 

\begin{equation}
h = -f_1\left(r\right) dt^2 + \frac{h_{1}\left(r\right)}{f_{1}\left(r\right)} dr^2 + r^2 d\Omega^2
\label{eqn:hansatz}
\end{equation}
for two arbitrary functions $f_1$ and $h_1$. Using the ansatz \eqref{eqn:hansatz} in \eqref{eqn:kappaSchwarz}, we obtain two independent equations - one for the $rr$ components and another for the $\theta\theta$ (proportional to the $\phi \phi$) components for the functions $f_1$ and $h_1$. Notice that we do not have the temporal equations for the $tt$ components. Taking the sum and difference of the numerators of the $rr$ and $\theta\theta$ equations, we obtain two equations for $f_1$ and $h_1$: one that is independent of $\kappa$ and another that depends on $\kappa$. It can then be shown that since both these equations must simultaneously hold, the only consistent solution is the Schwarzschild solution (where the terms proportional to $\kappa$ in the latter equation identically vanish). We thus see that $\kappa$ and $\bar{\kappa}$ do not affect the Schwarzschild solution even when $\alpha = 1$. 

For the Schwarzschild solution to be physically meaningful, it must be matched to the interior metric of an object and it is this matching that specifies the mass term in the Schwarzschild solution. The interior solution is given by the Tollman Oppenheimer Volkoff (or TOV) equations. These equations are solutions to: 
\begin{equation}
R_{ij}\left(1 - \bar{\kappa}\right) - \left(\frac{1}{4} +  \frac{\kappa}{2}\right) R h_{ij} = - \left( T_{ij} - \frac{1}{4}T h_{ij}\right)
\label{eqn:kappaSchwarz}
\end{equation}
whose solution requires specification of an equation of state for the matter. In weak gravity, the non-linear (gravitational) corrections from the TOV equations are negligible - the interior metric can be constructed using a convolution of the matter distribution with the Green's function of linearized gravity. But, this Green's function is in fact the Schwarzschild solution and as we have seen, it is independent of $\bar{\kappa}$ and $\kappa$. Thus, there are no constraints on these from experiments that are performed in weak gravity. 

This argument fails in strong gravity. Specifically, while the Schwarzschild solution is the unique solution that describes the exterior metric of a gravitating object even in strong gravity, the actual parameters of the solution are determined by how it is matched to the interior TOV solution. In strong gravity, corrections to the TOV metric that depend on $\kappa$ are possible. In fact, they exist - the correction to cosmological evolution caused by $\kappa$ is proof of its impact in strong gravity.  But, other than cosmology, we do not have experimental access to the interior metric of any other strongly gravitating system. Moreover, the existence of $\kappa$ is also degenerate with the equation of state of the body (as in the case of cosmology, where it is degenerate with the amount of dark matter) and the existence of such a correction to the General Relativistic effects on the interior are difficult to reliably extract.

\subsection{Cosmic Inflation}
\label{subsec:inflation}
In this sub-section, we discuss the seemingly paradoxical fact that our solution permits inflationary cosmology while explicitly de-gravitating the vacuum energy. The resolution to this ``paradox'' lies in the fact that while the vacuum energy of a field does not gravitate, the kinetic energy of a field does source gravitation. The equations of motion of the field depend upon the slope of its potential, but not its absolute magnitude. If a scalar field has a sufficiently shallow slope, its kinetic energy (cast in terms of proper time) can be nearly constant during the expansion of the universe and thus lead to an inflationary phase. For simplicity of analysis, consider a scalar field $\phi$ whose potential $V\left(\phi\right) = \Sigma \phi + V_0$, where $\Sigma$ is the slope of the linear potential for $\phi$ and $V_0$ is a constant. In regular gravity, $V_0$ couples to gravitation and is the source of the cosmological constant - but, it does not impact the equations of motion of $\phi$. The latter only depend on $\frac{\partial V}{\partial \phi} = \Sigma$. In our case, $V_0$ is irrelevant - it does not impact either the motion of $\phi$ or the gravitation of the scale factor $a\left(t\right)$. These equations are: 
\begin{equation}
2 \ddot{\phi} a^{\frac{3}{2}} + 3 \sqrt{a}\dot{a} \dot{\phi}+ \Sigma= 0 
\end{equation}
for the field $\phi$ and 
\begin{equation}
4 a \left(1 + 12 \kappa\right) \ddot{a} + \left(-1 + 12 \kappa\right) \dot{a}^2+ 8 a^2 \dot{\phi}^2 = 0
\end{equation}
for gravity. We have ignored the contribution from $\bar{\kappa}$, since unless we are in a superposition with different universes, as evident from the above analysis, $\bar{\kappa}$ is degenerate with redefining the values of $\kappa$ and $G_N$. It can now be checked that the late time solutions to this equation are of the form $a\left(t\right) = a_0 t^{4/3}$ and $\phi\left(t\right) = \phi_0 \log\left(t\right)$, leading to expansion with a constant Hubble parameter (as measured using proper time), resulting in inflation. The source of inflation in this case is not the vacuum energy $V_0$ - this term does not appear anywhere in the equations and the Hubble parameter during inflation is not set by it. Rather, the source of inflation is the kinetic energy $\dot{\phi} = \phi_0/t$, which when expressed in terms of proper time is  $\frac{d\phi}{d\tau} = \frac{dt}{d\tau} \frac{d\phi}{dt} = a\left(t\right)^{3/4} \frac{\phi_0}{t}$, which is independent of $t$ when $a\left(t\right) \propto t^{4/3}$. With $\kappa < -1/10$ as assumed, the shadow stress energy tensor from the initial conditions for $a, \dot{a}$ redshifts away - the dynamics is entirely determined by $\phi$. Ultimately, $\phi_0$ is set by the slope $\Sigma$ - this parameter controls the actual Hubble scale during inflation. 

An arbitrarily long slope can thus cause the universe to inflate for an arbitrarily long time. This long period of slow roll can be made to end by raising the potential at some point, at which point the field will oscillate around that minimum. The existence of an inflationary phase in this framework when there is a scalar field with a shallow slope also makes it possible for this framework to accommodate the observed existence of dark energy. Since the absolute value of the potential never gravitates, this framework suggests that the observed dark energy could be due to a quintessence field. 

However, as discussed in appendix \ref{Stability}, the simple model of inflation presented here is unstable when we consider superpositions of universes that are perturbatively different. This analysis showed that while scalar field models of inflation were unstable, accelerating universes that are perturbatively stable could be constructed from perfect fluid sources of energy density. Thus, to more robustly realize inflation in this framework, we would  need to realize inflation in such models. Alternately, one could also add non-linear quantum mechanical terms into the dynamics of the inflaton itself, causing perturbatively different initial inflaton conditions to evolve into a single coherent state at a rate faster than gravity. We do not pursue these issues in this paper. 

\section{Conclusions}
\label{sec:conclusions}

Quantum Mechanics and General Relativity are the two great pillars on which we have constructed physics in the last century. It is widely believed that both of these frameworks are resistant to modification. Conventional wisdom holds that while General Relativity is modified in the ultra-violet, its infra-red phenomenology cannot be altered. Quantum Mechanics is believed to be even more resistant to modification. Attempts to describe the ultra-violet nature of gravity are still constructed by assuming that the principles of linear quantum mechanics continue to hold. This wisdom persists even in light of astonishing experimental problems - the naive confluence of linear quantum mechanics and General Relativity predicts cosmic expansion that is at least 30 orders of magnitude faster than empirically observed. Discordance at such scale ought to provoke a re-examination of our fundamental principles, especially when such principles are axiomatic. 

In prior work,  we had shown that the linearity of quantum mechanics was not a logical requirement of a consistent quantum field theory. It is possible to construct theories with state dependent Hamiltonians that nevertheless describe a causal world with massless gauge interactions. We also discovered that these non-linear modifications can be $\mathcal{O}\left(1\right)$, without being experimentally constrained, despite over a century of experiments that have verified the predictions of linear quantum mechanics. The key reason for this astonishing fact is that many non-linear quantum mechanical phenomena are suppressed if our quantum state (the one that we can manipulate) has a suppressed overlap with the full quantum state of the entire universe. However, even in this case, non-linear effects on operators that have a non-zero expectation value in our vacuum, such as the metric and the higgs fields do persist. These fields are both separately associated with their own respective hierarchy problems that are difficult to solve using the tools of linear quantum field theory. Non-linear quantum mechanical effects however can alter the rules of conventional quantum field theory and provide new opportunities to tackle these problems. 

In this paper, we used such non-linear terms to modify the infra-red phenomenology of gravity - a task that has not proven to be difficult using linear quantum mechanics. This modification was specifically constructed to solve the cosmological constant problem in unimodular gravity. In unimodular gravity, the vacuum energy density of the standard model does not gravitate. But, in a quantum theory, the initial conditions of the universe need not obey the classical constraint equations of Einstein. The failure of the constraint appears as a shadow stress energy tensor in the effective classical description of the theory. In conventional unimodular gravity, this shadow stress energy tensor has the equation of state of the cosmological constant - in effect, it implies that the late time dynamics of the universe is governed by the initial conditions of the metric and its derivatives, as opposed to the matter density in it. The infra-red modification of gravity that we have constructed permits this initial condition to decay - that is we modify the equation of state of the shadow stress energy tensor so that this energy density redshifts away. The late time dynamics of the universe is governed by the matter and radiation in it, as experimentally observed. 

Even though this is  a large infrared modification to gravity, it is ill constrained. Precision tests of gravity have been performed in weakly gravitating local systems such as the solar system and local test masses.  Due to the weakness of gravity, these tests probe the vacuum metric around the object. Any modification of gravity in which the Schwarzschild solution is the unique solution to point masses is likely to be in concordance with these measurements. This is true in the modification proposed by us. We thus find that purely local tests of gravity do not constrain this scenario. 

Tests of gravity in the nonlinear regime are possible using cosmological measurements. If we are in case where $\alpha \ll 1$, it may be possible to search for $\bar{\kappa}$ by trying to extract  $G_N$  cosmologically and comparing it to local measurements of this quantity. This parameter is not strictly necessary for our solution - we just need  $\kappa$. However, the case for its existence is amplified by its ability to stabilize the dynamics of the quantum state when different universes interfere with each other. However, a direct test of  $\kappa$, the core of our solution,  is considerably harder. This is because $\kappa$ produces $\bar{R}$ - in all local measurements of gravity, where gravitation is weak, this quantity is vanishingly small. This term is also negligible during radiation domination. The only known time in the universe when it is relevant is during the eras of ``dark matter'' and ``dark energy'' domination where the unknown nature of both of these fluids makes it difficult to constrain $\kappa$. One possible avenue to search for $\kappa$ might be inflationary cosmology - if inflation did occur, $\bar{R}$ could be large and it could potentially lead to observable differences between the scalar and tensor spectrum.  This possibility deserves further study\footnote{S.R. is specifically grateful to Tony Padilla and David Stefanyszyn for discussions about inflation that lead to the realization that this model was consistent with inflation. }. We note that given the instability in simple inflationary models in this framework, further work is necessary to realize a complete model.   In this framework, the existence of dark energy is likely to be due to a quintessence field, whose slope is responsible for the accelerated expansion of the universe. The kinetic energy of this field could potentially be coupled to the Standard Model and it would be worth performing experiments to directly search for such an interaction \cite{Berghaus:2020ekh, Graham:2020kai, Carroll:1998zi, Pospelov:2004fj}. It is also likely that there is curvature in the slope of the field's potential. These features would cause the equation of state of dark energy to differ from -1, leading to potentially observable signatures in measurements of the equation of state of dark energy \cite{DESI:2025fii}. The ability of non-linear quantum mechanics to solve such outstanding problems of particle physics also motivates a broader experimental program to search for such effects in other quantum fields such as electromagnetism \cite{Broz:2022aea, Melnychuk:2024ngm,Polkovnikov:2022hkg}.




\section*{Acknowledgments}
We thank Tony Padilla, David Stefanyszyn and Raman Sundrum for useful discussions.   This work was supported by the U.S.~Department of Energy~(DOE), Office of Science, National Quantum Information Science Research Centers, Superconducting Quantum Materials and Systems Center~(SQMS) under Contract No.~DE-AC02-07CH11359. D.E.K.~and S.R.~are supported in part by the U.S.~National Science Foundation~(NSF) under Grant No.~PHY-1818899.
S.R.~is also supported by the Simons Investigator Grant No.~827042.  D.E.K.~is also supported by the Simons Investigator Grant No.~144924.

\begin{appendices}
\section{Perturbative Stability}
\label{Stability}
The perturbative stability of \eqref{eqn:ActualFix} can be analyzed through the following methods. To get a flavor of the analysis, we begin by considering the simple case where $|\Psi\rangle = \alpha |U\rangle + \beta |M\rangle$ where we take $|M\rangle$ to be a universe whose initial state is perturbatively different from that of $|U\rangle$. Assume that the scale factor in $|U\rangle$ is $a$ and the scale factor in $|M\rangle$ is $b$. The evolution of $|\Psi\rangle$ is governed by the following equations: 

\begin{eqnarray}
R^{a}_{ij} - \frac{1}{4} R^{a} h^{a}_{i j} -\frac{\kappa}{2} \sqrt{\frac{|g^{a}|}{\langle |g|\rangle}} \bar{R} h^{a}_{i j} - \bar{\kappa} \left(\frac{\langle |g|\rangle}{|g^{a}|}\right)^{\frac{1}{4}} \bar{R}_{ij} &=& -\left( T^{a}_{ij} - \frac{1}{4} T^{a} h^{a}_{ij}\right) \nonumber \\
R^{b}_{ij} - \frac{1}{4} R^{b} h^{b}_{i j} -\frac{\kappa}{2} \sqrt{\frac{|g^{b}|}{\langle |g|\rangle}} \bar{R} h^{b}_{i j} - \bar{\kappa} \left(\frac{\langle |g|\rangle}{|g^{b}|}\right)^{\frac{1}{4}} \bar{R}_{ij} &=& -\left( T^{b}_{ij} - \frac{1}{4} T^{b} h^{b}_{ij}\right) 
\label{eqn:Perturb}
\end{eqnarray}

where the upper index $a, b$ refers to the scale factors $a$, $b$ of $|U\rangle$ and $|M\rangle$ respectively, with quantities such as $R^{a,b}$ constructed for those universes. We analyze this system for stress tensors that are perfect fluids. In the perturbative case, both universes have stress energy tensors with the same equation of state $w$. In the absence of the perturbation, we know that the long term evolution of these states is $\propto t^{\frac{8}{3\left( 3 + w \right)}}$ (see discussion below \eqref{eqn:sourceexp}). To analyze the pertubative stability, we write $a\left(t\right) = \left( a_0 + \delta a_1\left(t\right)\right) t^{\frac{8}{3\left(3 + w\right)}}$ and $b\left(t\right) = \left( b_0 + \delta b_1\left(t\right)\right) t^{\frac{8}{3\left(3 + w\right)}}$. We substitute this ansatz into \eqref{eqn:Perturb} and expand the equations to linear order in $\delta a_1\left(t\right)$ and $\delta b_1\left(t\right)$. We then ask if the solutions to $\delta a_1$ and $\delta b_1$ decay in time (indicating stability) or grow in time (indicating instability). The algebraic expressions obtained by following this procedure are complex and we do not present them in this paper. Instead, we summarize the results of this analysis. The differential equations that we obtain by following this procedure are a set of linear homogeneous Cauchy-Euler equations of the form: 
\begin{equation}
t^2 M V'' + t N V' + O V = 0
\end{equation}
where the vector $V = \left(\delta a_1, \delta b_1\right)$ and the matrices $M, N$ and $O$ are complicated functions that depend on the parameters $\alpha, \beta, \kappa, \bar{\kappa}$ and the sources of the stress energy tensor $T^{a,b}$. They also depend on the points $\left(a_0, b_0\right)$ around which the solution is expanded. To determine the late time behavior of $\delta a_1$, $\delta b_1$, we try solutions of the form $\delta a_1 \propto t^{p_1}, \delta b_1 \propto t^{p_2}$. This converts the differential equation to an algebraic equation in $p_{1,2}$. If the real parts of the roots of this characteristic equation are all negative,  the solutions are stable and decay with time. If the real part of at least one root is positive, then there are unstable growing modes. 

In principle, one could try to find the functional criteria that would force all of the real parts of the roots of the characteristic equation to be negative. But, the equations are sufficiently complicated that we did not attempt to find such a criterion. Instead, we tried various random values of $\left(\alpha, \beta, \kappa, \bar{\kappa}\right)$ and find that for perfect fluids with $w > -1$, even with $\bar{\kappa}$ as small as $\sim -0.01$, for values of $\kappa$ of interest to us ($\sim \mathcal{O}\left(-1\right)$), it is possible to find values of $\left(a_0, b_0\right)$ for which all the real parts of the roots are negative, proving the existence of perturbatively stable solutions. It can be checked that such points do not exist when $\bar{\kappa} = 0$, necessitating the need for this term, the physical origin of which is tied to the need to effectively violate the null energy condition in the equations of motion so that the scale factors can bounce and oscillate as the perturbation damps out. 

Given that we have stable solutions when $|\Psi\rangle = \alpha |U\rangle + \beta |M\rangle$, it is easy to see that such stable solutions also exist when $|\Psi\rangle = \alpha |U\rangle + \sum_{i} \beta_i |M_i\rangle$, with $|M_i\rangle$ perturbatively close to $|U\rangle$. The proof of their existence comes from the following observation - start with the fact that when all the $|M_i\rangle$ are equal to $|M\rangle$, the situation is identical to that of the case when we just had two states $|U\rangle$ and $|M\rangle$. Thus, we see that $|\Psi\rangle = \alpha |U\rangle + \sum_{i} \beta_i |M\rangle$ is stable under perturbations. Now perturb one of the $|M_i\rangle$ away from $|M\rangle$. The characteristic roots of the Cauchy-Euler equations are continuous functions of these parameters - thus if they were initially negative, for small perturbations they will continue to be negative. Stability is thus guaranteed. 

\begin{figure}[h!]
    \centering
    \begin{minipage}{0.48\textwidth}
        \centering
        \includegraphics[width=\textwidth]{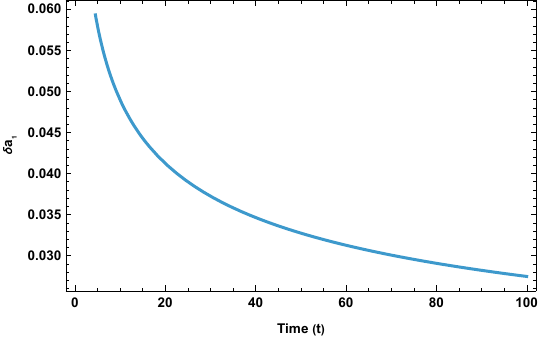}
    \end{minipage}\hfill
    \begin{minipage}{0.48\textwidth}
        \centering
        \includegraphics[width=\textwidth]{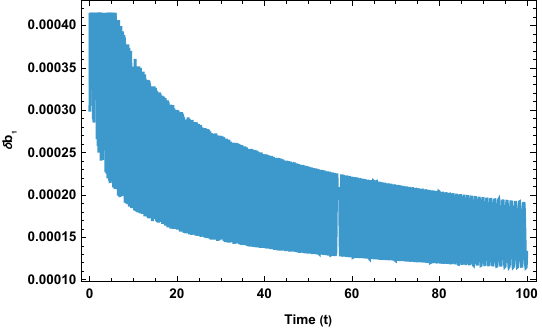}
    \end{minipage}
    \caption{The analytic solution for $w = 0$ with $\kappa = -2$, $\bar{\kappa} = -0.025$. The solutions are expanded around $a_0 \approxeq 0.84$ and $b_0 \approxeq 0.0025$ with $\alpha = 1/3.$}
    \label{fig:analyticw0}
\end{figure}

\begin{figure}[h!]
    \centering
    \begin{minipage}{0.48\textwidth}
        \centering
        \includegraphics[width=\textwidth]{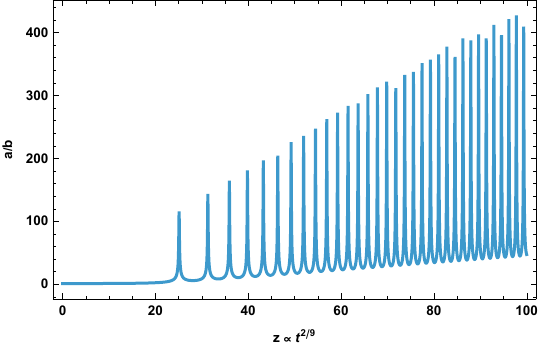}
    \end{minipage}\hfill
    \begin{minipage}{0.48\textwidth}
        \centering
        \includegraphics[width=\textwidth]{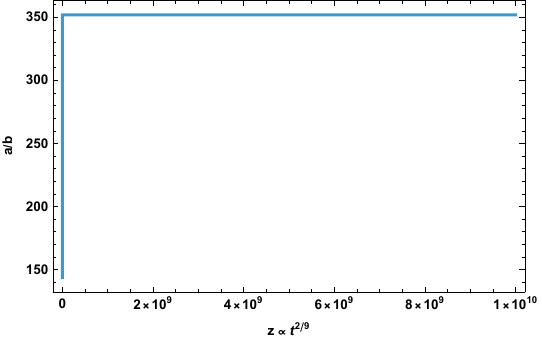}
    \end{minipage}
    \caption{The numerical solution for $w = 0$ with $\alpha = 1/3$, $\kappa = -2$, $\bar{\kappa} = -0.025$ plotted as the ratio of the scale factors $a/b$. The initial conditions for the metric and matter are perturbed around 10 percent. Instead of unimodular time, we show the redshift $z$ that the solution captures. Note that the ratio a/b that the solution asymptotes to is the ratio of $a_0/b_0$ around which the analytic solution is expanded in figure \ref{fig:analyticw0}. }
    \label{fig:numericw0}
\end{figure}

\begin{figure}[h!]
    \centering
    \begin{minipage}{0.48\textwidth}
        \centering
        \includegraphics[width=\textwidth]{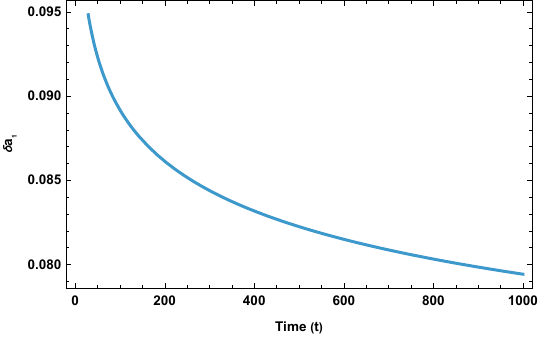}
    \end{minipage}\hfill
    \begin{minipage}{0.48\textwidth}
        \centering
        \includegraphics[width=\textwidth]{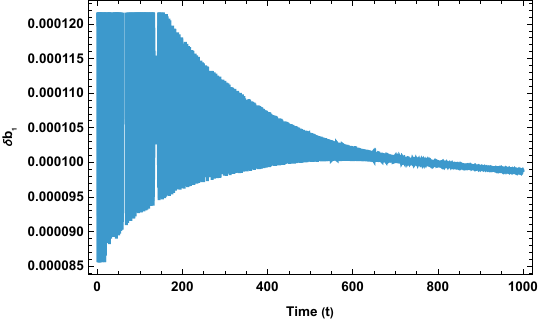}
    \end{minipage}
    \caption{The analytic solution for $w = 1/3$ with $\kappa = -0.5$, $\bar{\kappa} = -0.025$. The solutions are expanded around $a_0 \approxeq 2.33$ and $b_0 \approxeq 0.001$ with $\alpha = 1/10.$}
    \label{fig:analyticwr}
\end{figure}

\begin{figure}[h!]
    \centering
    \begin{minipage}{0.48\textwidth}
        \centering
        \includegraphics[width=\textwidth]{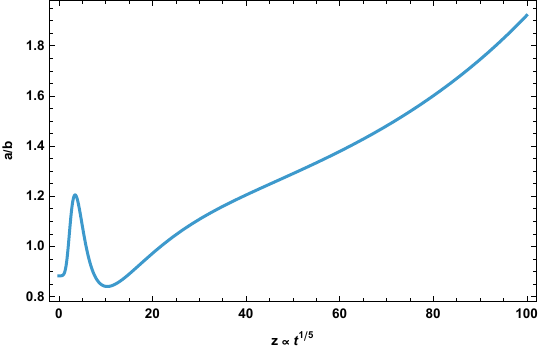}
    \end{minipage}\hfill
    \begin{minipage}{0.48\textwidth}
        \centering
        \includegraphics[width=\textwidth]{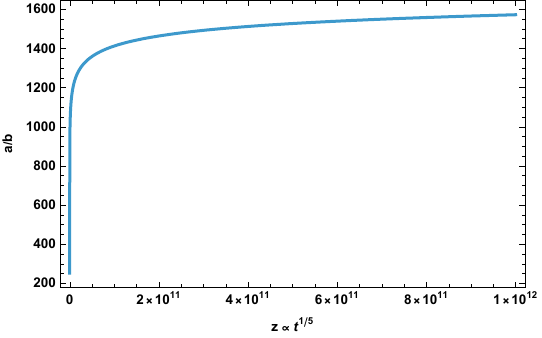}
    \end{minipage}
    \caption{The numerical solution for $w = 1/3$ with $\alpha = 1/10$, $\kappa = -1/2$, $\bar{\kappa} = -0.025$ plotted as the ratio of the scale factors $a/b$. The initial conditions for the metric and matter are perturbed around 10 percent. Instead of unimodular time, we show the redshift $z$ that the solution captures. The ratio $a/b$ slowly asymptotes to the stable ratio around which the analytic solution in figure \ref{fig:analyticwr} is expanded around.}
    \label{fig:numericwr}
\end{figure}

\begin{figure}[h!]
    \centering
    \begin{minipage}{0.48\textwidth}
        \centering
        \includegraphics[width=\textwidth]{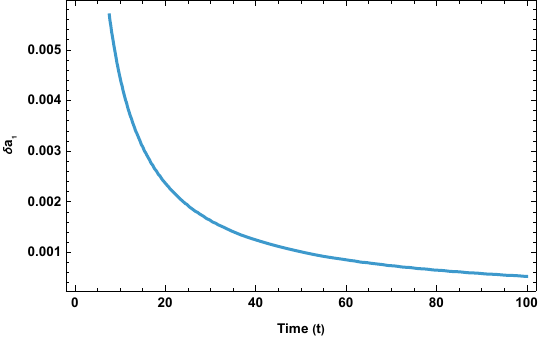}
    \end{minipage}\hfill
    \begin{minipage}{0.48\textwidth}
        \centering
        \includegraphics[width=\textwidth]{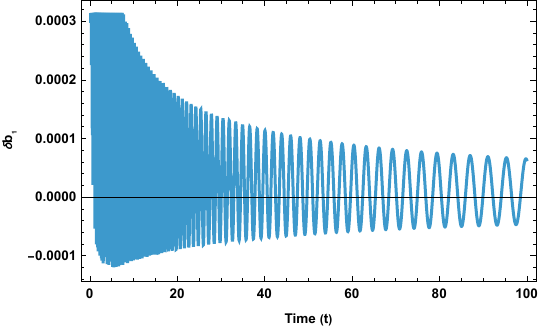}
    \end{minipage}
    \caption{The analytic solution for $w = -0.9$ with $\kappa = -2$, $\bar{\kappa} = -0.025$. The solutions are expanded around $a_0 \approxeq 0.053$ and $b_0 \approxeq 0.001$ with $\alpha = 1/3.$}
    \label{fig:analyticwA}
\end{figure}

\begin{figure}[h!]
    \centering
    \begin{minipage}{0.48\textwidth}
        \centering
        \includegraphics[width=\textwidth]{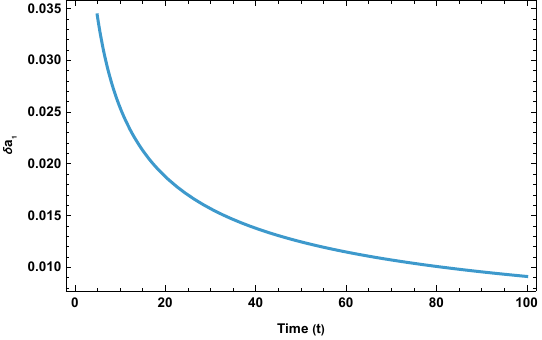}
    \end{minipage}\hfill
    \begin{minipage}{0.48\textwidth}
        \centering
        \includegraphics[width=\textwidth]{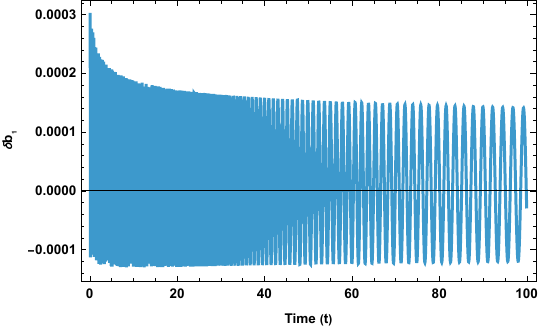}
    \end{minipage}
    \caption{The analytic solution for $w = 1/3$ with $\kappa = -1/5$, $\bar{\kappa} = -0.01$. The solutions are expanded around $a_0 \approxeq 2.23$ and $b_0 \approxeq 0.0033$ with $\alpha = 1/3.$}
    \label{fig:analyticwrnc}
\end{figure}

\begin{figure}[h!]
    \centering
    \begin{minipage}{0.48\textwidth}
        \centering
        \includegraphics[width=\textwidth]{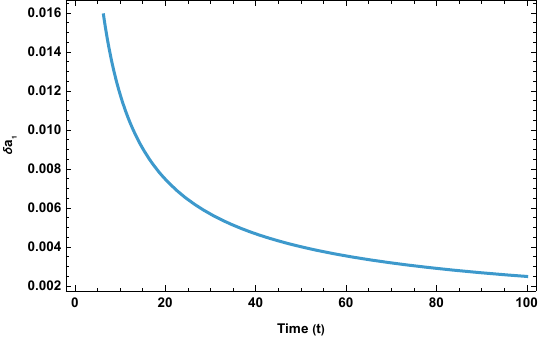}
    \end{minipage}\hfill
    \begin{minipage}{0.48\textwidth}
        \centering
        \includegraphics[width=\textwidth]{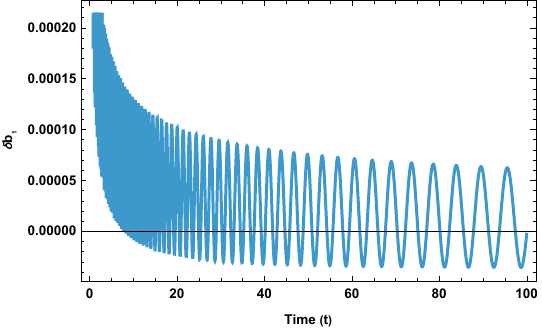}
    \end{minipage}
    \caption{The analytic solution for $w = 0$ with $\kappa = -1/5$, $\bar{\kappa} = -0.01$. The solutions are expanded around $a_0 \approxeq 1.75$ and $b_0 \approxeq 0.0097$ with $\alpha = 1/3.$}
    \label{fig:analyticw0nc}
\end{figure}

The solutions to this analytic analysis is shown for three cases: $w = 0$ (matter domination, figure \ref{fig:analyticw0}), $w = 1/3$ (radiation domination, figure \ref{fig:analyticwr}) and a perfect fluid source with $w = -0.9$ (corresponding to a source that produces accelerated expansion, figure \ref{fig:analyticwA}). We also show the numerical solutions to \eqref{eqn:Perturb} in figures \ref{fig:numericw0} (matter domination) and \ref{fig:numericwr}. We see that the full numerical solution with a perturbative  ($\sim 0.1$) deviation in initial conditions evolves to a stable evolution, with the ratio of scale factors $a/b$ settling to the ratio set by the stable region set by the analytic analysis. If the initial perturbation is too large, it is also possible for the system to not approach this stable solution and end up in a runaway situation. In plotting the numerical solution, instead of plotting the unimodular time, we plot the redshift $z$ over which the evolution occurs since this is a physical measure of time.  The match between the analytic and numerical solutions provides confidence in the numerical solution. It can also be checked that if $\bar{\kappa}$ is set to zero, the numerical solution rapidly becomes unstable, further confirming the match between the analytic and numerical analysis. In these plots we have taken $\bar{\kappa} = -0.025$ - this is largely motivated by the uncertainty on the actual limit on $\bar{\kappa}$ due to the Hubble tension. Solutions similar to the ones plotted exist even when $\bar{\kappa} = -0.01$, which is well below the limit on $\bar{\kappa}$ when the Hubble tension is ignored (see figures \ref{fig:analyticw0nc}, \ref{fig:analyticwrnc}).  

\begin{figure}[h!]
    \centering
    \begin{minipage}{0.48\textwidth}
        \centering
        \includegraphics[width=\textwidth]{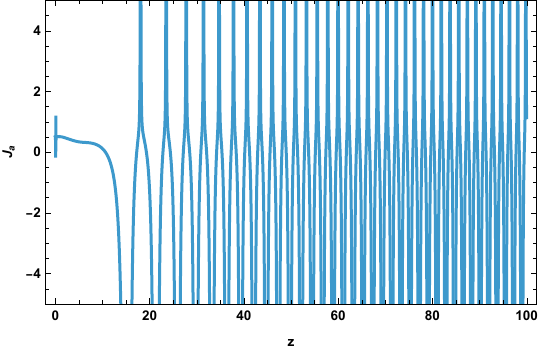}
    \end{minipage}\hfill
    \begin{minipage}{0.48\textwidth}
        \centering
        \includegraphics[width=\textwidth]{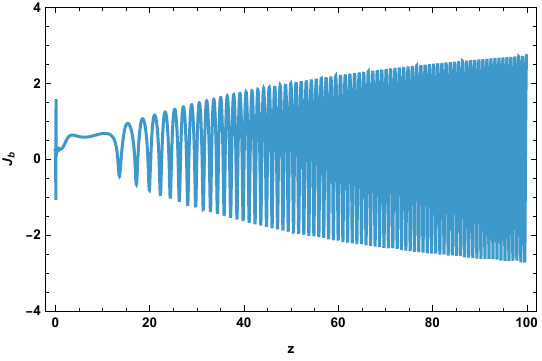}
    \end{minipage}
    \caption{The numerical solution for two universes with two different fluids. The universe with scale factor $a$ is matter dominated and the universe with scale factor $b$ is radiation dominated. While the solutions evolve stably, the effective acceleration $J_{a,b}$ is not what would be expected in a universe dominated by either radiation or matter. The interference causes the effective equation of state to oscillate. As a reference, for matter domination $J_a = 5/8$, $J_b = 1/2$. The quantity $J$ is defined as $J = \frac{K}{H^2} a\left(t\right)$ where $H = a'\left(t\right) a\left(t\right)^{3/4}$, $K = a\left(t\right)^{3/4}\frac{d}{dt}\left(a\left(t\right)^{3/4} \frac{da}{dt}\right)$. For a universe undergoing a simple power law expansion, this quantity is time independent and can be mapped to the equation of state driving the power law expansion simply from the cosmological evolution of the scale factor.}
    \label{fig:numerictwofluids}
\end{figure}

\begin{figure}[h!]
    \centering
    \begin{minipage}{0.48\textwidth}
        \centering
        \includegraphics[width=\textwidth]{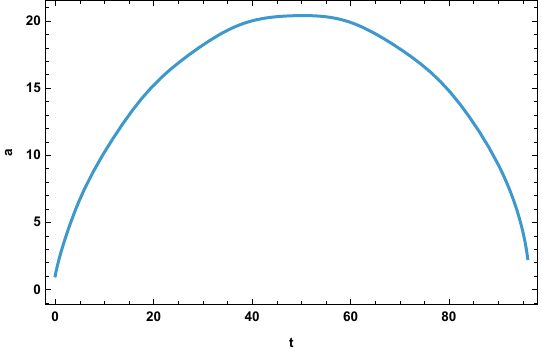}
    \end{minipage}\hfill
    \begin{minipage}{0.48\textwidth}
        \centering
        \includegraphics[width=\textwidth]{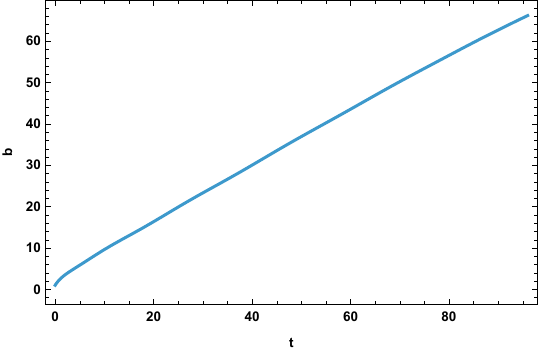}
    \end{minipage}
    \caption{Unstable solution for slow roll inflation where the interference of the two universe causes one to contract and the other to expand.}
    \label{fig:inflation}
\end{figure}

The confidence in the numerical analysis implies that it can be trusted when it is difficult to perform an analytic stability analysis. This situation arises when we consider quantum states of the form $|\Psi\rangle = \alpha |U\rangle + \beta |M\rangle$ where the two states are very different. For example, we can consider the state $|U\rangle$ to be matter dominated and the state $|M\rangle$ to be radiation dominated. Even in this case, the numerical solution shows that the two interfering universes evolve in a stable fashion (at least for some time, see figure \ref{fig:numerictwofluids}). However, the effective cosmological equation of state that they observe in their respective universes will not look like that of pure radiation or matter domination - instead, the effective equation of state oscillates. 

This analysis was also performed for the case of slow roll inflation a potential of the form $V =  \frac{1}{2}m^2 \phi^2$.  In this case, both the analytic analysis and the numerical computation (see figure \ref{fig:inflation}) show the existence of an unstable growing mode for any non-zero value of $\kappa$, irrespective of the value of $\bar{\kappa}$. Thus, if inflation is to be realized in this scenario in a stable way when different universes could interact with each other, one would need a different model to obtain the required accelerated expansion. For example, we do find that a perfect fluid model with $w = -0.9$ can in fact be in a perturbatively stable superposition. For the case of $V\left(\phi\right) = \frac{1}{2} m^2 \phi^2$, we also  numerically checked to ensure that the solution is perturbatively stable when we are in the fast roll regime. This is as expected since in the fast roll regime, the scalar field oscillates about its minimum and thus acts as a perfect fluid with equation of state $w = 0$. The answer is thus consistent with the perfect fluid analysis performed for matter.

\section{Mode Analysis}
\label{Modes}

The gravitational quantum theory constructed by us consists of six opertors $g_{ij}$ and their conjugate momenta $\pi^{ij}$. In this section, we show that the theory still only contains two propagating degrees of freedom, corresponding to the massless tensor modes of the graviton. The evolution of the other four degrees of freedom are determined by the matter sources and the initial conditions of the metric ({\it i.e.} the shadow matter stress energy tensor). The absence of new massless degrees of freedom implies that corrections to the UV cutoff of the theory due to the $\kappa$, $\bar{\kappa}$ terms will remain analytic in $\kappa$ and $\bar{\kappa}$. The mode analysis can be performed by looking for solutions to the vacuum field equations of the theory in the limit of linearized gravity. That is, we need solutions to: 

\begin{equation}
R_{ij} - \frac{1}{4} R h_{ij} - \frac{\kappa}{2} \sqrt{\frac{|g|}{\langle |g|\rangle}} \bar{R} h_{ij} - \bar{\kappa} \left(\frac{\langle |g|\rangle}{|g|} \right)^{\frac{1}{4}}\bar{R}_{ij} = 0
\end{equation}

An important aspect of analyzing these equations is the nature of $\bar{R}$ and $\bar{R}_{ij}$ in comparison to the terms $R$ and $R_{ij}$ in these equations. In linear quantum mechanics, the propagation of particles ({\it i.e.} Fock states) can be obtained simply by solving the classical field equations. In fact, the classical field equations are identities that are obeyed by the expectation values of field operators. This equation is non-trivial for coherent states and can produce useful information - but for fock space states, since the expectation values of various field operators is zero, these equations trivially hold and do not produce useful information. However, in linear quantum mechanics, due to linearity, the time evolution of coherent states (classical waves) can be mapped to the time evolution of particle states. When we are interested in understanding the existence of additional propagating modes in a theory, we are interested in the propagation of Fock space states - these are precisely the states that appear in perturbation theory. In non-linear quantum mechanics, since the evolution is in fact state dependent, the propagation of fock states and coherent states are in fact different. 

In the vacuum, $\bar{R}$ and $\bar{R}_{ij}$ are in fact zero. Thus, when we are interested in computing the propagation of Fock states, the vacuum equations of interest are identical to that of General Relativity. Thus there are no new propagating modes. It is also interesting to ask if there is even the possibility of new propagating modes for coherent states. For any non-trivial effects from the new terms, we need $\bar{R}$ and $\bar{R}_{ij}$ to be non-zero. At the linearized level, this then implies that the coherent state equations are of the form: 

\begin{equation}
R_{ij}\left(1 - \bar{\kappa}\right) - \left(\frac{1}{4} + \frac{\kappa}{2}\right) R \eta_{ij} = 0
\end{equation}
where $\eta_{ij}$ is the metric of flat space.  By explicitly solving these equations at the linearized level, one can show that  wave-like solutions only exist for  the transverse, traceless modes - {\it i.e.} gravitational waves. Further since we are happy with $\bar{\kappa}$ as small as $\sim -0.03$, there is no concern about these parameters changing the signs of the kinetic terms of the gravitons. It can be checked that $\kappa$ is irrelevant for this analysis.

\end{appendices}

\bibliographystyle{unsrt}
\bibliography{references}

@article{SupernovaSearchTeam:1998fmf,
    author = "Riess, Adam G. and others",
    collaboration = "Supernova Search Team",
    title = "{Observational evidence from supernovae for an accelerating universe and a cosmological constant}",
    eprint = "astro-ph/9805201",
    archivePrefix = "arXiv",
    doi = "10.1086/300499",
    journal = "Astron. J.",
    volume = "116",
    pages = "1009--1038",
    year = "1998"
}

@article{Kaloper:2023kua,
    author = "Kaloper, Nemanja",
    title = "{Axion flux monodromy discharges relax the cosmological constant}",
    eprint = "2307.10365",
    archivePrefix = "arXiv",
    primaryClass = "hep-th",
    doi = "10.1088/1475-7516/2023/11/032",
    journal = "JCAP",
    volume = "11",
    pages = "032",
    year = "2023"
}

@article{Kaloper:2022jpv,
    author = "Kaloper, Nemanja and Westphal, Alexander",
    title = "{Quantum-mechanical mechanism for reducing the cosmological constant}",
    eprint = "2204.13124",
    archivePrefix = "arXiv",
    primaryClass = "hep-th",
    reportNumber = "DESY 22-070",
    doi = "10.1103/PhysRevD.106.L101701",
    journal = "Phys. Rev. D",
    volume = "106",
    number = "10",
    pages = "L101701",
    year = "2022"
}

@article{Kaloper:2013zca,
    author = "Kaloper, Nemanja and Padilla, Antonio",
    title = "{Sequestering the Standard Model Vacuum Energy}",
    eprint = "1309.6562",
    archivePrefix = "arXiv",
    primaryClass = "hep-th",
    doi = "10.1103/PhysRevLett.112.091304",
    journal = "Phys. Rev. Lett.",
    volume = "112",
    number = "9",
    pages = "091304",
    year = "2014"
}

@article{Weinberg:1988cp,
    author = "Weinberg, Steven",
    editor = "Hsu, Jong-Ping and Fine, D.",
    title = "{The Cosmological Constant Problem}",
    reportNumber = "UTTG-12-88",
    doi = "10.1103/RevModPhys.61.1",
    journal = "Rev. Mod. Phys.",
    volume = "61",
    pages = "1--23",
    year = "1989"
}

@article{Page:1981aj,
    author = "Page, Don N. and Geilker, C. D.",
    title = "{Indirect Evidence for Quantum Gravity}",
    reportNumber = "PRINT-81-0221 (PENN-STATE)",
    doi = "10.1103/PhysRevLett.47.979",
    journal = "Phys. Rev. Lett.",
    volume = "47",
    pages = "979--982",
    year = "1981"
}

@article{Linde:2008xf,
    author = "Linde, Andrei D. and Vanchurin, Vitaly and Winitzki, Sergei",
    title = "{Stationary Measure in the Multiverse}",
    eprint = "0812.0005",
    archivePrefix = "arXiv",
    primaryClass = "hep-th",
    doi = "10.1088/1475-7516/2009/01/031",
    journal = "JCAP",
    volume = "01",
    pages = "031",
    year = "2009"
}

@article{Kaplan:2005rr,
    author = "Kaplan, David E. and Sundrum, Raman",
    title = "{A Symmetry for the cosmological constant}",
    eprint = "hep-th/0505265",
    archivePrefix = "arXiv",
    doi = "10.1088/1126-6708/2006/07/042",
    journal = "JHEP",
    volume = "07",
    pages = "042",
    year = "2006"
}

@article{Abbott:1984qf,
    author = "Abbott, L. F.",
    title = "{A Mechanism for Reducing the Value of the Cosmological Constant}",
    reportNumber = "BRX-TH-175",
    doi = "10.1016/0370-2693(85)90459-9",
    journal = "Phys. Lett. B",
    volume = "150",
    pages = "427--430",
    year = "1985"
}

@article{Graham:2017hfr,
    author = "Graham, Peter W. and Kaplan, David E. and Rajendran, Surjeet",
    title = "{Born again universe}",
    eprint = "1709.01999",
    archivePrefix = "arXiv",
    primaryClass = "hep-th",
    doi = "10.1103/PhysRevD.97.044003",
    journal = "Phys. Rev. D",
    volume = "97",
    number = "4",
    pages = "044003",
    year = "2018"
}

@article{Graham:2019bfu,
    author = "Graham, Peter W. and Kaplan, David E. and Rajendran, Surjeet",
    title = "{Relaxation of the Cosmological Constant}",
    eprint = "1902.06793",
    archivePrefix = "arXiv",
    primaryClass = "hep-ph",
    doi = "10.1103/PhysRevD.100.015048",
    journal = "Phys. Rev. D",
    volume = "100",
    number = "1",
    pages = "015048",
    year = "2019"
}

@article{Arkani-Hamed:2000hpr,
    author = "Arkani-Hamed, Nima and Dimopoulos, Savas and Kaloper, Nemanja and Sundrum, Raman",
    title = "{A Small cosmological constant from a large extra dimension}",
    eprint = "hep-th/0001197",
    archivePrefix = "arXiv",
    reportNumber = "SU-ITP-00-04",
    doi = "10.1016/S0370-2693(00)00359-2",
    journal = "Phys. Lett. B",
    volume = "480",
    pages = "193--199",
    year = "2000"
}

@article{Kachru:2000hf,
    author = "Kachru, Shamit and Schulz, Michael B. and Silverstein, Eva",
    title = "{Selftuning flat domain walls in 5-D gravity and string theory}",
    eprint = "hep-th/0001206",
    archivePrefix = "arXiv",
    reportNumber = "SLAC-PUB-8337, SU-ITP-00-02, IASSNS-HEP-00-05",
    doi = "10.1103/PhysRevD.62.045021",
    journal = "Phys. Rev. D",
    volume = "62",
    pages = "045021",
    year = "2000"
}

@article{Kaplan:2021qpv,
    author = "Kaplan, David E. and Rajendran, Surjeet",
    title = "{Causal framework for nonlinear quantum mechanics}",
    eprint = "2106.10576",
    archivePrefix = "arXiv",
    primaryClass = "hep-th",
    doi = "10.1103/PhysRevD.105.055002",
    journal = "Phys. Rev. D",
    volume = "105",
    number = "5",
    pages = "055002",
    year = "2022"
}

@article{Polchinski:1990py,
    author = "Polchinski, Joseph",
    title = "{Weinberg's nonlinear quantum mechanics and the EPR paradox}",
    reportNumber = "NSF-ITP-90-101, UTTG-21-90",
    doi = "10.1103/PhysRevLett.66.397",
    journal = "Phys. Rev. Lett.",
    volume = "66",
    pages = "397--400",
    year = "1991"
}

@article{Burns:2022fzs,
    author = "Burns, Anne-Katherine and Kaplan, David E. and Melia, Tom and Rajendran, Surjeet",
    title = "{Time Evolution in Quantum Cosmology}",
    eprint = "2204.03043",
    archivePrefix = "arXiv",
    primaryClass = "gr-qc",
    reportNumber = "FERMILAB-PUB-22-949-SQMS-V",
    month = "4",
    year = "2022"
}

@article{Kaplan:2023wyw,
    author = "Kaplan, David E. and Melia, Tom and Rajendran, Surjeet",
    title = "{The Classical Equations of Motion of Quantized Gauge Theories, Part I: General Relativity}",
    eprint = "2305.01798",
    archivePrefix = "arXiv",
    primaryClass = "hep-th",
    reportNumber = "FERMILAB-PUB-23-227-SQMS-V",
    month = "5",
    year = "2023"
}

@article{Kaplan:2023fbl,
    author = "Kaplan, David E. and Melia, Tom and Rajendran, Surjeet",
    title = "{The Classical Equations of Motion of Quantized Gauge Theories, Part 2: Electromagnetism}",
    eprint = "2307.09475",
    archivePrefix = "arXiv",
    primaryClass = "hep-th",
    reportNumber = "FERMILAB-PUB-23-391-SQMS-V",
    month = "7",
    year = "2023"
}

@article{Melnychuk:2024ngm,
    author = "Melnychuk, Oleksandr and others",
    title = "{Improved bound on nonlinear quantum mechanics using a cryogenic radio frequency experiment}",
    eprint = "2411.09611",
    archivePrefix = "arXiv",
    primaryClass = "quant-ph",
    reportNumber = "FERMILAB-PUB-24-0725-SQMS-T-TD",
    doi = "10.1103/gkg6-fqsc",
    journal = "Phys. Rev. D",
    volume = "112",
    number = "1",
    pages = "012020",
    year = "2025"
}

@article{Polkovnikov:2022hkg,
    author = "Polkovnikov, Mark and Gramolin, Alexander V. and Kaplan, David E. and Rajendran, Surjeet and Sushkov, Alexander O.",
    title = "{Experimental Limit on Nonlinear State-Dependent Terms in Quantum Theory}",
    eprint = "2204.11875",
    archivePrefix = "arXiv",
    primaryClass = "quant-ph",
    reportNumber = "FERMILAB-PUB-22-977-SQMS-V",
    doi = "10.1103/PhysRevLett.130.040202",
    journal = "Phys. Rev. Lett.",
    volume = "130",
    number = "4",
    pages = "040202",
    year = "2023"
}

@article{Broz:2022aea,
    author = "Broz, Joseph and You, Bingran and Khan, Sumanta and Haeffner, Hartmut and Kaplan, David E. and Rajendran, Surjeet",
    title = "{Test of Causal Nonlinear Quantum Mechanics by Ramsey Interferometry with a Trapped Ion}",
    eprint = "2206.12976",
    archivePrefix = "arXiv",
    primaryClass = "quant-ph",
    doi = "10.1103/PhysRevLett.130.200201",
    journal = "Phys. Rev. Lett.",
    volume = "130",
    number = "20",
    pages = "200201",
    year = "2023"
}

@article{Graham:2020kai,
    author = {Graham, Peter W. and Hac{\i}{\"o}mero{\u{g}}lu, Selcuk and Kaplan, David E. and Omarov, Zhanibek and Rajendran, Surjeet and Semertzidis, Yannis K.},
    title = "{Storage ring probes of dark matter and dark energy}",
    eprint = "2005.11867",
    archivePrefix = "arXiv",
    primaryClass = "hep-ph",
    doi = "10.1103/PhysRevD.103.055010",
    journal = "Phys. Rev. D",
    volume = "103",
    number = "5",
    pages = "055010",
    year = "2021"
}

@article{Berghaus:2020ekh,
    author = "Berghaus, Kim V. and Graham, Peter W. and Kaplan, David E. and Moore, Guy D. and Rajendran, Surjeet",
    title = "{Dark energy radiation}",
    eprint = "2012.10549",
    archivePrefix = "arXiv",
    primaryClass = "hep-ph",
    doi = "10.1103/PhysRevD.104.083520",
    journal = "Phys. Rev. D",
    volume = "104",
    number = "8",
    pages = "083520",
    year = "2021"
}

@article{Carroll:1998zi,
    author = "Carroll, Sean M.",
    title = "{Quintessence and the rest of the world}",
    eprint = "astro-ph/9806099",
    archivePrefix = "arXiv",
    reportNumber = "NSF-ITP-98-063",
    doi = "10.1103/PhysRevLett.81.3067",
    journal = "Phys. Rev. Lett.",
    volume = "81",
    pages = "3067--3070",
    year = "1998"
}

@article{Pospelov:2004fj,
    author = "Pospelov, M. and Romalis, M.",
    title = "{Lorentz invariance on trial}",
    doi = "10.1063/1.1784301",
    journal = "Phys. Today",
    volume = "57N7",
    pages = "40--46",
    year = "2004"
}

@article{Kaplan:2025zjj,
    author = "Kaplan, David E. and Rajendran, Surjeet",
    title = "{A Solution to the Hierarchy Problem with Non-Linear Quantum Mechanics}",
    eprint = "2510.12030",
    archivePrefix = "arXiv",
    primaryClass = "hep-ph",
    month = "10",
    year = "2025"
}

@article{Lamine:2024xno,
    author = "Lamine, Brahim and Ozdalkiran, Yacob and Mirouze, Louis and Erdogan, Furkan and Ilic, St{\'e}phane and Tutusaus, Isaac and Kou, Raphael and Blanchard, Alain",
    title = "{Cosmological measurement of the gravitational constant G using the CMB, BAO, and BBN}",
    eprint = "2407.15553",
    archivePrefix = "arXiv",
    primaryClass = "astro-ph.CO",
    doi = "10.1051/0004-6361/202451602",
    journal = "Astron. Astrophys.",
    volume = "697",
    pages = "A109",
    year = "2025"
}

@article{Benevento_2022,
   title={An Exploration of an Early Gravity Transition in Light of Cosmological Tensions},
   volume={935},
   ISSN={1538-4357},
   url={http://dx.doi.org/10.3847/1538-4357/ac80fd},
   DOI={10.3847/1538-4357/ac80fd},
   number={2},
   journal={The Astrophysical Journal},
   publisher={American Astronomical Society},
   author={Benevento, Giampaolo and Kable, Joshua A. and Addison, Graeme E. and Bennett, Charles L.},
   year={2022},
   month=aug, pages={156} }

@article{DESI:2025fii,
    author = "Lodha, K. and others",
    collaboration = "DESI",
    title = "{Extended Dark Energy analysis using DESI DR2 BAO measurements}",
    eprint = "2503.14743",
    archivePrefix = "arXiv",
    primaryClass = "astro-ph.CO",
    reportNumber = "FERMILAB-PUB-25-0164-PPD",
    month = "3",
    year = "2025"
}

@article{Weinberg:1987dv,
    author = "Weinberg, Steven",
    title = "{Anthropic Bound on the Cosmological Constant}",
    reportNumber = "UTTG-06-87",
    doi = "10.1103/PhysRevLett.59.2607",
    journal = "Phys. Rev. Lett.",
    volume = "59",
    pages = "2607",
    year = "1987"
}

@article{Brown:1987dd,
    author = "Brown, J. David and Teitelboim, C.",
    title = "{Dynamical Neutralization of the Cosmological Constant}",
    doi = "10.1016/0370-2693(87)91190-7",
    journal = "Phys. Lett. B",
    volume = "195",
    pages = "177--182",
    year = "1987"
}

@article{Brown:1988kg,
    author = "Brown, J. David and Teitelboim, C.",
    title = "{Neutralization of the Cosmological Constant by Membrane Creation}",
    doi = "10.1016/0550-3213(88)90559-7",
    journal = "Nucl. Phys. B",
    volume = "297",
    pages = "787--836",
    year = "1988"
}

@article{Sundrum:2003tb,
    author = "Sundrum, Raman",
    title = "{Fat euclidean gravity with small cosmological constant}",
    eprint = "hep-th/0310251",
    archivePrefix = "arXiv",
    doi = "10.1016/j.nuclphysb.2004.05.011",
    journal = "Nucl. Phys. B",
    volume = "690",
    pages = "302--330",
    year = "2004"
}

\end{document}